\documentclass[twocolumn]{aastex63}
\pdfoutput=1 

\usepackage{amsmath,amstext}
\usepackage[T1]{fontenc}
\usepackage{ar} 
\usepackage[figure,figure*]{hypcap}
\usepackage{bm}

\newcommand{\x}{{\bf{x}}}
\renewcommand{\t}{{\bm{\theta}}}
\newcommand{\xobs}{{\bf{x}_\text{obs}}}
\newcommand{\tobs}{{\bm{\theta}_\text{obs}}}
\newcommand{\xsim}{{\bf{x}_\text{sim}}}
\newcommand{\tsim}{{\bm{\theta}_\text{sim}}}

\newcommand{\sobs}{{\bm{\sigma}_\text{obs}}}

\newcommand{\uniform}[2]{Uniform from {#1} to {#2}}
\newcommand*{\figuretitle}[1]{%
    {\centering
    \textbf{#1}
    \par\medskip}
}

\defcitealias{Wolfgang2016}{W16}
\defcitealias{LUVOIR2019}{L19}

\shorttitle{Identifying Exo-Earth Candidates in Direct Imaging Data}
\shortauthors{Bixel \& Apai}

\begin{document}

\title{Identifying Exo-Earth Candidates in Direct Imaging Data Through Bayesian Classification}

\author{Alex Bixel}
\affiliation{Department of Astronomy/Steward Observatory, The University of Arizona, 933 N. Cherry Avenue, Tucson, AZ 85721, USA}
\affiliation{Earths in Other Solar Systems Team, NASA Nexus for Exoplanet System Science}

\author{D\'aniel Apai}
\affiliation{Department of Astronomy/Steward Observatory, The University of Arizona, 933 N. Cherry Avenue, Tucson, AZ 85721, USA}
\affiliation{Earths in Other Solar Systems Team, NASA Nexus for Exoplanet System Science}
\affiliation{Lunar and Planetary Laboratory, The University of Arizona, 1640 E. University Blvd, AZ 85721, USA}

\begin{abstract}
Future space telescopes may be able to directly image $\sim$10 -- 100 planets with sizes and orbits consistent with habitable surface conditions (``exo-Earth candidates'' or EECs), but observers will face difficulty in distinguishing these from the potentially hundreds of non-habitable ``false positives'' which will also be detected. To maximize the efficiency of follow-up observations, a prioritization scheme must be developed to determine which planets are most likely to be EECs. In this paper, we present a Bayesian method for estimating the likelihood that any directly imaged extrasolar planet is a true exo-Earth candidate by interpreting the planet's apparent magnitude and separation in light of existing exoplanet statistics. As a specific application of this general framework, we use published estimates of the discovery yield of future space-based direct imaging mission concepts to conduct ``mock surveys'' in which we compute the likelihood that each detected planet is an EEC. We find that it will be difficult to determine which planets are EECs with $>50\%$ confidence using single-band photometry immediately upon their detection. The best way to reduce this ambiguity would be to constrain the planet's orbit by revisiting the system multiple times or through a radial velocity precursor survey. Astrometric or radial velocity constraints on the planet's mass would offer a lesser benefit. Finally, we show that a Bayesian approach to prioritizing targets would improve the follow-up efficiency of a direct imaging survey versus a blind approach using the same data. For example, the prioritized approach could reduce the amount of integration time required for the spectral detection (or rejection) of water absorption in most EECs by a factor of two.
\end{abstract}

\keywords{astrobiology, space telescopes, Bayesian statistics, astronomical techniques, habitable planets, direct imaging}

\section{Introduction} \label{sec:introduction}
One of the primary science goals driving the development of new space telescopes is the detection and characterization of Earth-like planets around nearby stars. Spectroscopy of the planets' reflected light spectra would reveal whether they are potentially habitable -- i.e., liquid water could exist on their surfaces. The presence of biosignature gasses such as O$_2$, O$_3$, and CH$_4$ could be interpreted as evidence for life beyond the Earth \citep[e.g.,][]{Seager2016,Fujii2018}, although this interpretation would not be straightforward as many of these gasses can be produced abiotically \citep[e.g.,][]{Catling2018,Meadows2018}.

Recently, final study reports have been published for two space telescope design concepts with a primary science goal of directly imaging and spectroscopically characterizing Earth-like planets around nearby stars. These are the Habitable Exoplanet Observatory \citep[HabEx,][]{HabEx2019} and the Large UV/Optical/Infrared Surveyor \citep[LUVOIR,][hereafter \citetalias{LUVOIR2019}]{LUVOIR2019}. Following \cite{Stark2014}, both reports provide estimates for the expected yield of planets across a range of sizes and insolations. The HabEx report predicts the detection and characterization of $12^{+18}_{-8}$ approximately Earth-sized planets in the habitable zone with a 4-meter aperture, while the LUVOIR report predicts $51^{+75}_{-33}$ with the 15-meter aperture ``LUVOIR-A'' design. These design concepts have been thoroughly investigated, and it is likely that the design of any future direct imaging space telescope would enable it to detect comparable numbers of potentially habitable planets with moderate S/N photometry.

\begin{figure*}[p]
    \centering
    \figuretitle{\large False positives for exo-Earth candidate detections}
    \includegraphics[width=\textwidth]{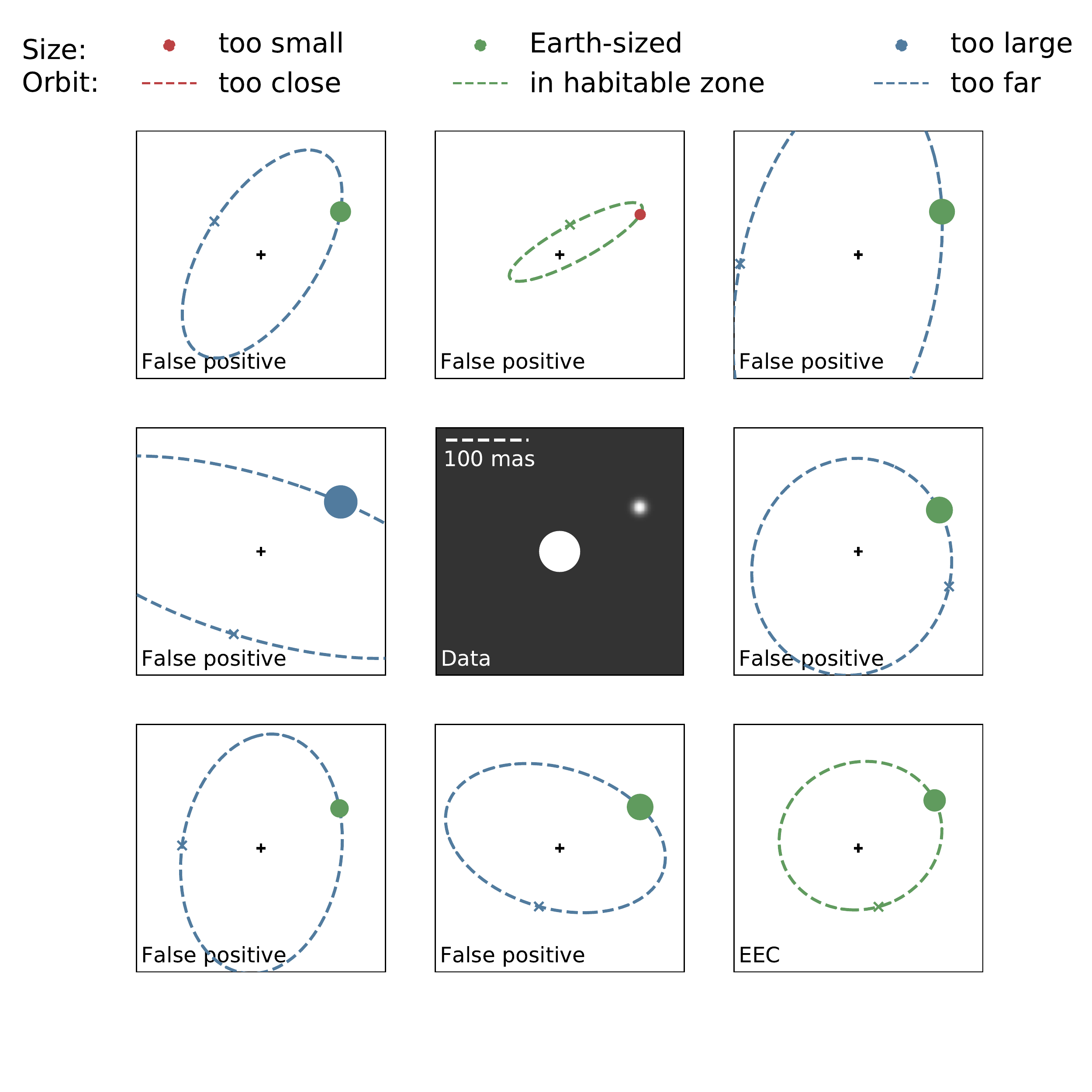}
    \caption{To illustrate the degeneracies which affect the interpretation of direct imaging data, we simulate the detection of a planet orbiting a Solar-type star at 15 parsecs (center panel), as well as several planets of varying sizes, orbits, and albedos which have a similar projected separation and magnitude (surrounding panels). It is not clear whether this data point represents a true exo-Earth candidate, or one of many potential false positives. The color and size of each circle represents the planet's potential radius; only green points are approximately Earth-sized ($\sim 0.8 - 1.6 \,R_\oplus$). The color of the potential orbit represents its insolation; only green orbits are in the habitable zone. An `x' marks the planet's closest approach to the observer.}
    \label{fig:gallery}
\end{figure*}

However, these dozens of ``exo-Earth candidates'' would be detected amidst hundreds of planets with atmospheric compositions or equilibrium temperatures not conducive to Earth-like life, including planets outside of the habitable zone, large mini-Neptunes with thick H/He envelopes, or Mars-sized planets which have been stripped of their atmospheres. These non-habitable planets often demonstrate the same \emph{observable} parameters (e.g., apparent magnitude and separation) as the exo-Earth candidates, but are far more common and may therefore cause a significant number of false positive detections. We demonstrate this problem in Figure \ref{fig:gallery}. In fact, \cite{Guimond2018} show that separation- and contrast-based selection criteria could suffer from a false discovery rate as high as 77\% given just the detection data in a single band, or 47\% if prior constraints on the orbit are available.

While the ``false positive'' planets would be interesting to characterize in their own right, spectroscopy across the full wavelength range could take \emph{weeks} for the faintest targets, diverting time and resources from higher priority targets. To separate the potentially habitable and non-habitable planets, a survey could use multi-epoch broadband photometry to characterize their orbits and spectroscopic observations of H$_2$O absorption features to confirm the presence of water vapor in habitable zone targets. However, even within the habitable zone, there exist many planets that are too small or large to be habitable, and the identification of water absorption features will require a significant investment of time \citep{Kawashima2019}. Additionally, warmer planets within the runaway greenhouse limit could exhibit water absorption features. In order to maximize the efficiency of an exo-Earth imaging mission, it is necessary to develop a method for identifying those planets that are most likely to be Earth analogs using the limited data which will be available upon detection.

We have previously advocated for a Bayesian approach to assessing the potential habitability of newly detected exoplanets \citep{Apai2018}. The Bayesian approach allows one to probabilistically constrain the properties of the planet by leveraging knowledge from exoplanet statistics on planet radii, masses, and orbital properties. It also allows one to fold in predictions from theoretical models of planet formation and evolution as prior knowledge. As an example, in \cite{Bixel2017} we used the Monte Carlo method to infer the likely composition of Proxima Cen b \citep{Anglada-Escude2016} in light of well-established statistical priors and the limited data available about the system. We found that it is $\sim 90\%$ likely that the planet is small and rocky as opposed to a ``mini-Neptune''. In this paper, we extend this approach to assess the likelihood that a directly imaged planet has an appropriate composition and orbit to be potentially habitable.

In Section \ref{sec:framework} we review our Bayesian framework and give an example of how it can be applied to characterize directly imaged planets. In Section \ref{sec:priors} we discuss the prior assumptions upon which our framework is based, and how they might be improved in the coming decade. Using the planet yield estimates provided in LUVOIR final report as a baseline estimate for the yield of a hypothetical direct imaging mission, in Section \ref{sec:mock_survey} we conduct mock surveys where we detect each target and estimate the apparent likelihood that it has a potentially habitable composition and orbit. In Section \ref{sec:results} we discuss the results of these surveys, including what types of non-habitable planets would be mistaken for potentially habitable planets, and which additional data could help to resolve this ambiguity. Finally, we show that our approach to target prioritization could greatly enhance the efficiency of follow-up observations after all of the planets have been detected.

\section{A Bayesian Framework for Classifying Directly Imaged Planets} \label{sec:framework}
\subsection{Monte Carlo method} \label{sec:monte_carlo}
Here we review the Monte Carlo method for Bayesian inference. This method allows an observer to constrain the \emph{unobservable} properties of a planet based on limited precision measurements of its observable data values by assuming some understanding of the prior distribution of intrinsic properties and their relationship to the observable data values.

Suppose planets can be described by some set of intrinsic properties $\t$ and some resulting set of observable data values $\x$ which can be calculated from $\t$. Given a prior probability distribution for the values of $\t$, $P(\t)$, then the probability distribution for $\t$ \emph{given} $\x$ can be calculated using Bayes' equation:
\[
P(\t|\x) = \frac{P(\x|\t)P(\t)}{\int_\theta P(\x|\t)P(\t)}
\]
The left term is commonly referred to as the posterior distribution of $\t$, and $P(\x|\t)$ as the likelihood function.

For most astrophysical applications there is no analytical solution to this equation, so it must be solved numerically. Under the Monte Carlo method, we use the prior probability distribution $P(\t)$ to simulate a set of properties $\tsim$, then calculate the simulated data values $\xsim$ directly. Next, we accept or reject this simulated planet based on the value of its likelihood function for some observed set of data values $\xobs$. Assuming the data values are drawn from independent, normal probability distributions with standard deviations (measurement uncertainties) $\sobs$, the likelihood function is that of a multivariate Gaussian:

\[
P(\x|\t) = \prod_{i=1}^m \exp((\xobs_{,i}-\xsim_{,i})^2/2\sobs_{,i}^2)
\]

If only an upper limit is available for a component of $\xobs$ (e.g., magnitude), then the prior sample is first pruned of simulated members exceeding that limit.

This procedure is repeated in parallel for a large number of planets until a statistically sufficient number are accepted. The result of the likelihood-based selection is a sample of planets whose properties $\tsim$ are distributed according to the posterior distribution. In other words, a histogram of the accepted values of $\tsim$ represents the probability distribution for $\tobs$, the properties of the observed planet.

\subsection{Constructing the prior sample} \label{sec:prior_sample}

\begin{deluxetable*}{p{1.5in}p{2.5in}p{2in}}
\tablecaption{A list of the prior assumptions which we use to build our sample in Case 1, along with relevant literature references. The additional cases in Table \ref{tab:cases} may modify these assumptions to reflect new prior knowledge or data.
\label{tab:priors}}
\tablehead{Parameter & Description of prior & Reference}
\startdata
Radius and period & \emph{Kepler} occurrence rates & \cite{Mulders2018} \\
Class & ``sub-terrestrial'', ``terrestrial'', or ``ice giant'' based on radius, following Figure \ref{fig:classes} & \cite{Fulton2017,Zahnle2017} \\
Mass & Empirical mass-radius relationship with intrinsic spread & \cite{Wolfgang2016} \\
Habitable zone boundaries & Planet mass-dependent LWHZ models (runaway and maximum greenhouse limits) & \cite{Kopparapu2014} \\
Albedo & \uniform{0.2}{0.7} &  \\
Eccentricity & Beta distribution ($\alpha = 0.867$, $\beta = 3.03$) & \cite{Kipping2013} \\
$\omega$, $\Omega$, M \tablenotemark{$\dagger$} & \uniform{0}{$2\pi$} & \\
$\cos(i)$ & \uniform{-1}{1} &  \\
Exo-Earth Candidates (EECs) & ``terrestrial'' class planets in the LWHZ & \\
\enddata
\tablenotetext{\dagger}{Argument of pericenter; longitude of the ascending node; mean anomaly}
\end{deluxetable*}

\begin{deluxetable*}{lp{6in}}
\tablecaption{The different cases under which we conduct our mock surveys, representing the different data which the observer might have. \label{tab:cases}}
\tablehead{Case & Description}
\startdata
1 & Each planet has a monochromatic geometric albedo drawn uniformly from 0.2 to 0.7. The planet's monochromatic magnitude and separation angle are observed in a single epoch with a signal-to-noise ratio of 7 and centroid precision $\sigma_c$ = 3.8 mas. \\ \hline
2 & Prior radial velocity (RV) observations provide constraints of $\pm 10\%$ on the period, $\pm 5$ cm/s on the radial velocity semi-amplitude, and $\pm 30^\circ$ on the mean anomaly, in addition to the data from Case 1. \\ \hline
3 & Simultaneous observations of a debris disk provide a $\pm 1^\circ$ constraint on the inclination and longitude of the ascending node of the orbital plane. The planet's orbital elements may be further offset from these by $\pm 0.2$ in $\Omega$ and $\cos(i)$. \\ \hline
4 & Multiple epochs of direct imaging data permit constraints of $\pm 15^\circ$ on the phase angle, $\pm 10\%$ on the semi-major axis, and $\pm 0.05$ on the eccentricity.. \\ \hline
5 & Case 4 with an additional $\pm 0.1 \mu\text{as}$ constraint on the semi-amplitude of the star's motion due to the planet. \\ \hline
6 & We assume prior knowledge about the albedo distribution of small planets. Each planet's spectral albedo is determined by its class and location with respect to the habitable zone. Random monochromatic and polychromatic offsets are also introduced. The planet's magnitude in three bands and separation angle are observed in a single epoch with a wavelength-integrated signal-to-noise ratio of 7. \\ \hline
7 & Combination of Cases 5 and 6: three-band photometry is available along with constraints on the planet's orbit and the star's astrometric motion from multi-epoch observations. \\ \hline
\enddata
\end{deluxetable*}

\begin{figure*}[p]
    \centering
    \figuretitle{\large Simulating the prior sample of planets}
    \includegraphics[width=0.9\textwidth]{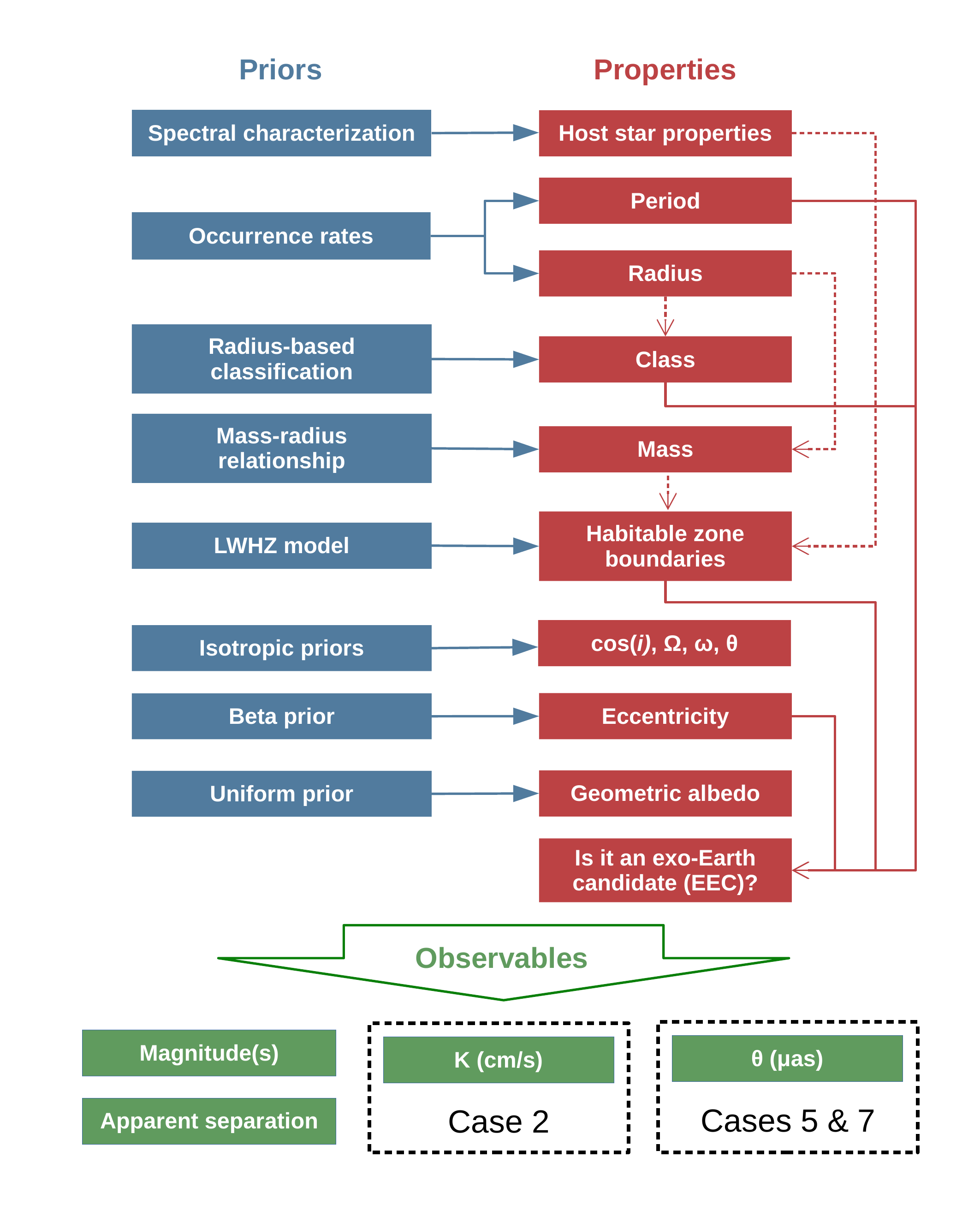}
    \caption{Flowchart illustrating our algorithm for simulating the prior sample of planets under Case 1. The red boxes represent the intrinsic planet properties which we simulate, and the red arrows indicate how they are used in the calculation of other properties. The blue boxes represent the priors which we use to simulate the properties, and the green boxes are the observable values which can be compared to the data.}
    \label{fig:flowchart1}
\end{figure*}

We use the priors in Table \ref{tab:priors} to construct the prior sample according to the algorithm visualized in Figure \ref{fig:flowchart1}, and we discuss the prior assumptions in detail in Section \ref{sec:priors}. We consider several cases governing the amount of data available to the observer - these are listed in Table \ref{tab:cases}.

Before we simulate the properties of the directly imaged planet, we first consider the properties of its host star, which the observer will only know with finite precision. We represent this measurement uncertainty by drawing a unique stellar radius and mass for each system from normal distributions with $\sigma$ = 3\% and 7\%. We assume nearly exact measurements of the distance and luminosity of the host star, and we discuss these assumptions in Section \ref{sec:stellar_properties}.

We generate the radius and period of the planet using \emph{Kepler} occurrence rates as a prior probability distribution, then use the radius to classify the planet as too small (``sub-terrestrial''), too large (``ice giant''), or of the proper size to maintain a habitable atmosphere against stellar irradiation (``terrestrial''). This requires us to extrapolate to planets smaller ($R \lesssim 0.5\, R_\oplus$) or with longer periods ($P \gtrsim 100$ days) than those readily available in the \emph{Kepler} sample, as we discuss in Section \ref{sec:occurrence_rates}. We calculate the mass from an empirical mass-radius relationship, where we assume some intrinsic variance due to stochastic planet formation histories and differences between the host stars of the planets on which these relationships are based. We draw eccentricities from a beta distribution and the remaining orbital elements are assumed to be isotropically distributed.

The liquid water habitable zone (hereafter LWHZ) is the range of orbital separations over which a broadly Earth-like planet could feasibly host liquid water on its surface. \cite{Kopparapu2014} find the zone's boundaries to be a function of the planet's mass and the star's effective temperature, hence we cannot infer a planet's membership to the LWHZ based solely on its insolation. We calculate the LWHZ boundaries and determine whether the planet lies in the runaway greenhouse, temperate, or maximum greenhouse regimes, interpolating between the discrete planet masses modeled by \cite{Kopparapu2014} (0.1, 1.0, and 5.0 $M_\oplus$), and taking the minimum or maximum mass values for planets outside of this range. For planets with non-circular orbits we use the mean flux approximation to determine whether they are in the habitable zone, which \cite{Bolmont2016} find to be valid for planets with low or moderate eccentricities receiving Earth-like insolations.

Finally, we assign a geometric albedo to each planet using one of two methods for each of the cases in Table \ref{tab:cases}. In Cases 1-5, we assign a monochromatic albedo from a broad uniform prior with no dependence on the planet's class or orbital parameters. In Cases 6 and 7, we generate a spectrum for each planet by mapping it to a solar system analog with a comparable size and orbit. These two cases allow us to test the usefulness of color information for identifying potential exo-Earths. We dedicate more considerable discussion to the geometric albedos in Section \ref{sec:types}.

\subsection{Calculating the observable data values} \label{sec:observables}
Once the full assortment of planet properties has been simulated, we can proceed to calculate the observable data values to compare against those of a newly detected planet.

\subsubsection{Apparent separation}

The angular separation vector has two components, and can be calculated from the orbital elements and the distance to the system $d$. We adopt the same reference frame and notation as \cite{Murray2010}\footnote{Figure 4 and Equations 53 \& 54}, where $i = 90^\circ$ is an ``edge-on'' inclination and the observer is at $z = \infty$, so the angular separation components orthogonal to the line of sight are:

\[
\begin{aligned}
s_x &= (r/d) [\cos(\Omega)\cos(\omega+f) - \sin(\Omega)\sin(\omega+f)\cos(i)]\\
s_y &= (r/d) [\sin(\Omega)\cos(\omega+f) + \cos(\Omega)\sin(\omega+f)\cos(i)]
\end{aligned}
\]

In most cases, the position angle has no meaningfully defined zero point, so for simplicity we calculate only the net separation $|s| = \sqrt{s_x^2+s_y^2}$. The exception is Case 4, where the coeval detection of a debris disk is used to constrain the orbital plane of the planet; therefore the position angle is meaningfully defined, and we compute $s_x$ and $s_y$ separately.

\subsubsection{Apparent magnitude(s)}

Following \cite{Madhusudhan2012}\footnote{Figure 1 and Equations 4 \& 33}, we model the planet as a Lambertian sphere, in which case the planet-to-star contrast ratio when observed at orbital phase $\alpha$ is:

\[
\frac{L(\lambda)}{L_*(\lambda)} = A_g \left(\frac{R_P}{a}\right)^2 \left[ \frac{\sin(\alpha)+(\pi-\alpha)\cos(\alpha)}{\pi} \right]
\]

The phase angle is $\alpha = \text{Cos}^{-1}[\sin(\omega + f)\sin(i)]$. We draw the geometric albedo from a prior distribution, and allow slightly super-Lambertian values ($A_g > 2/3$) as these are observed in some wavelength ranges in the solar system.

\subsubsection{Radial velocity and astrometric semi-amplitudes and periods}

The semi-amplitude of the star's periodic radial velocity variation, assuming no other perturbers and $M_P \ll M_*$, is:

\[
K = (8.95 \, \text{cm/s}) \frac{(M_P/M_\oplus)\sin(i)}{(M_*/M_\odot)^{1/2}(a/\text{AU})^{1/2}(1-e^2)^{1/2}}
\]

The semi-amplitude of the star's astrometric motion is:

\[
\theta = (3.00\, \mu\text{as}) \frac{(M_P/M_\oplus)}{(M_*/M_\odot)}\frac{(a/\text{AU})}{(d/\text{pc})}
\]

In both cases the period of the stellar motion (and therefore the planet's orbit) is also measurable. However the astrometric mass measurement requires that the system be observed for multiple epochs, in which case the planet's orbital period can be derived from its apparent motion about the host star.

\subsection{Calculating the posterior probability distributions} \label{sec:posterior_sample}
We generate a posterior sample of simulated planets following the scheme in Section \ref{sec:monte_carlo}, where $\tsim$ are the simulated properties from Section \ref{sec:prior_sample}, $\xsim$ are the simulated data values from Section \ref{sec:observables}, and $\xobs$ are the data values for the observed planet. This allows us to calculate a posterior probability distribution for each component of $\tobs$, the physical properties of the observed planet. Based on just the planet's apparent separation and magnitude, we are therefore able to place informative constraints on its semi-major axis, radius, and mass.

The posterior probability in which we are most interested is the probability that the planet is potentially habitable - i.e., a ``terrestrial''-class planet ($0.8 \lesssim R_P \lesssim 1.6$) within the LWHZ. We designate these planets as ``exo-Earth candidates'' (EECs), following the terminology of \cite{LUVOIR2019} and \cite{HabEx2019}, albeit including a slightly different range of sizes. Planets outside of this range we designate as ``false positives'', with the \emph{false positive probability} being the likelihood that the observed planet - as judged based solely on its observed data values - is a false positive instead of an EEC.

\subsection{Example: an exo-Earth candidate around a Solar twin}

\begin{figure*}[t]
    \centering
    \includegraphics[width=\textwidth]{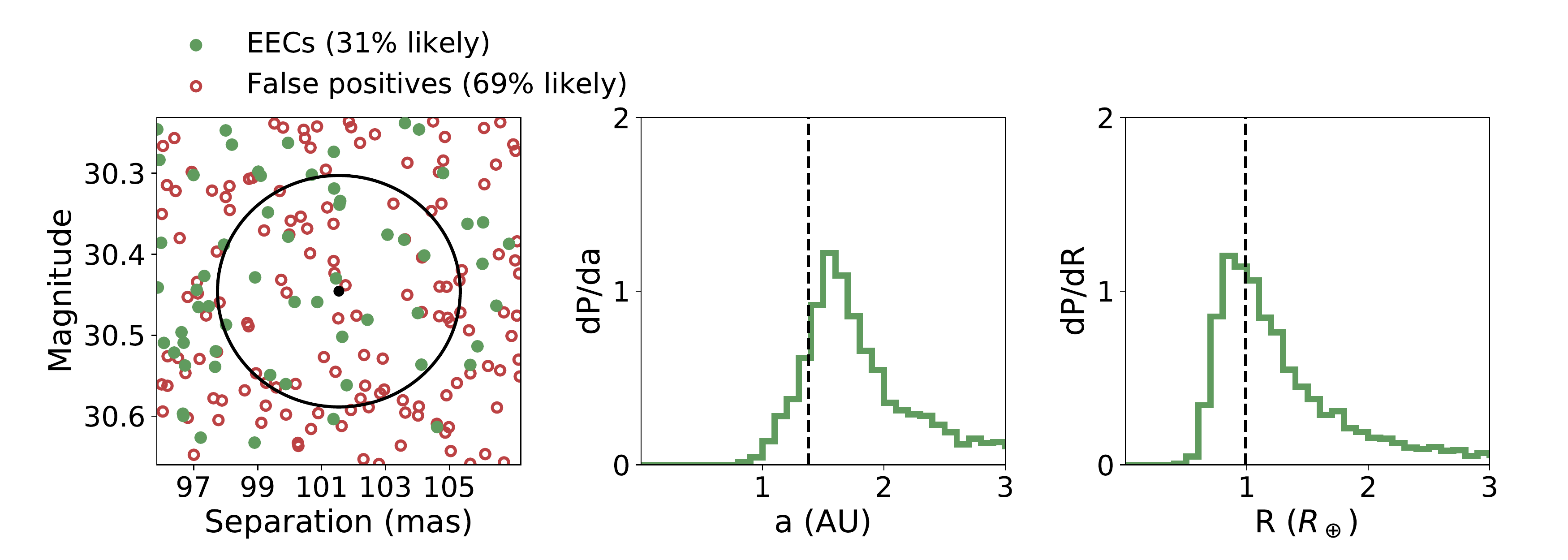}
    \caption{We simulate the detection of an ``ideal'' exo-Earth candidate from 15 pc. (Left) The separation- and magnitude- phase space populated by a host of simulated EECs (green) and non-EECs (red). The data point with uncertainties is marked in black. (Middle/right) Posterior distributions for the planet's semi-major axis and radius, taking into account \emph{Kepler} statistics and other priors. The true values are marked with dashed lines; this planet appears to the observer to have a wider orbit than it actually does.}
    \label{fig:example}
\end{figure*}

As a demonstration of our method, we use the procedure in Section \ref{sec:prior_sample} to generate an Earth-sized planet in the center of the LWHZ of a Solar-type star at 15 parsecs - an ideal exo-Earth candidate. We simulate the detection of this planet by assuming a S/N = 7 measurement of its monochromatic magnitude, and a $\pm 3.8$ mas measurement of its centroid.

Acting as the observer - who has no prior knowledge about the planet's true size and orbit - we use the procedure outlined above to estimate the likelihood that that the planet is an exo-Earth candidate based on its observed magnitude and separation. The results of our analysis are summarized in Figure \ref{fig:example}. We can confidently say that the planet is at least 1 AU from its star, and is unlikely to be farther than 3 AU - however, we cannot constrain its orbit to the habitable zone with certainty. We can also tell that the planet is almost certainly larger than $0.5 \, R_\oplus$ and smaller than $3\, R_\oplus$ - but this range includes planets which are too small (sub-terrestrial) or too large (sub-Neptune) to be EECs.

Most notably, we determine that it is only $31\%$ likely that this planet is an EEC, as the majority of simulated planets which have a similar separation and magnitude are not habitable. In this specific example, even though the planet appears $70\%$ likely to have a \emph{size} consistent with habitability, it also appears $60\%$ likely to orbit beyond the maximum greenhouse limit. This example shows that while it will be difficult to discriminate between true EECs and false positives given just the data available on detection, it will still be possible to place meaningful probabilistic constraints on the planet's properties.

\section{Prior assumptions} \label{sec:priors}
In this section, we discuss our priors in more detail by reviewing the relevant literature and discussing how they may be improved upon by future observations and modeling efforts.

\subsection{Stellar properties} \label{sec:stellar_properties}
It is likely that much effort will be dedicated to characterizing the stellar targets of a direct imaging mission in advance of its launch. Still, the stellar properties will only be constrained with finite precision - potentially several percent - so it is important that our prior sample includes host stars spanning the range defined by the relevant uncertainties.

\emph{Gaia} DR2 \citep{Gaia2018} has already provided high precision ($10-100$ $\mu$as) parallax measurements for nearby F-M spectral type stars, so we treat this uncertainty as negligible. Optical/IR interferometry has allowed for the measurement of stellar radii to $\sim 3\%$ for targets at $\sim 10 - 100$ pc \citep[e.g.][]{Ligi2016}. Masses are more difficult to measure, so stellar atmosphere models are often used - as an example, \cite{Sharma2018} constrain model-dependent masses for more than 10,000 stars using high resolution spectroscopy, with a median precision of 7\%. Following these examples, we draw the radius and mass of the host star from normal distributions with widths of 3\% and 7\%, respectively.

\subsection{Radius and period} \label{sec:occurrence_rates}
\emph{Kepler} allowed for the precise calculation of planet occurrence for planets with periods shorter than 100 days. \cite{Mulders2018} find that the Kepler occurrence rates are well-described by independent broken power laws in both radius and period:

\[
\frac{dN_\text{pl}}{d\log{P}d\log{R}} \propto f_R(R)f_P(P)
\]

\noindent where

\begin{equation*}
\begin{split}
f_P(P) &= (P/P_\text{break})^{a_P} \text{\quad($P$ < $P_\text{break}$)}  \\
       &= (P/P_\text{break})^{b_P} \text{\quad($P$ $\geq$ $P_\text{break}$)}
\end{split}
\end{equation*}

\[
\begin{split}
f_R(R) &= (R/R_\text{break})^{a_R} \text{\quad($R$ < $R_\text{break}$)}  \\
       &= (R/R_\text{break})^{b_R} \text{\quad($R$ $\geq$ $R_\text{break}$)}
\end{split}
\]

The best-fit parameters for this model are  $(P_\text{break},a_P,b_P) = (12,1.5,0.3)$ and $(R_\text{break},a_R,b_R) = (3.3,-0.5,-6)$. The number of planets per system, $N_\text{pl}$, is found to be $\sim 5$. Multiplicity will be an important factor for direct imaging surveys, as it could confuse the interpretation of the data or allow for simultaneous follow-up observations of multiple planets - but the topic is outside of the scope of this work. Here, we treat each detected planet independently, and normalize the above power laws so that $N_\text{pl} = 1$.

To properly simulate the abundance of planets in the habitable zone of F, G, and K spectral type stars - as well as smaller or cooler planets which might be mistaken for them - requires us to extrapolate \emph{Kepler} occurrence rates beyond the range of parameters within which they are well-understood ($R \gtrsim 0.5\, R_\oplus$, $P \lesssim 100$ days). This extrapolation could be problematic; for example, the results of \cite{Chen2016} suggest that planets on wider orbits can maintain thick volatile envelopes better against hydrodynamical escape, so we might find an over-abundance of large planets on wide orbits. We can gain some insight by studying the dependence of planet radii on insolation around low-mass stars, but these results cannot necessarily be extrapolated to Solar-mass regimes.

Furthermore, while \emph{Kepler} was generally not sensitive to planets smaller than 0.5 $R_\oplus$, it is likely that some such planets will be found by direct imaging missions and could masquerade as exo-Earth candidates. It is therefore necessary that we extrapolate the power law of \cite{Mulders2018} down to 0.1 $R_\oplus$ to ensure that the potentially large number of Mercury-sized objects are represented in our simulations. However, since our cutoff for exo-Earth candidates is $R \approx 0.8\, R_\oplus$, planets smaller than 0.5 $R_\oplus$ are less likely to be mistaken for EECs, so this extrapolation should not substantially influence our results.

\subsection{Planet classes}
We employ a radius-based classification scheme to separate potentially habitable planets from those that are too small or too large to be habitable. This classification is motivated by two physical considerations affecting whether a planet can maintain a habitable atmosphere against irradiation over several Gyr.

Empirical evidence suggests a change in planet compositions between $1.5-2\, R_\oplus$. Multiple authors find evidence for a split in planet densities in this range, with planets larger than $\sim 1.5\, R_\oplus$ mostly having densities much lower than the Earth's \citep[e.g.,][]{Weiss2014,Rogers2015,Chen2017}. \cite{Fulton2017} find a relative lack of \emph{Kepler} planets with $R \sim 1.75\, R_\oplus$ compared to smaller or larger radii; this ``photoevaporation valley'' was predicted by several authors who show that smaller planets would lose thick envelopes due to hydrodynamic escape \citep[e.g.,][]{Owen2013}. We interpret both results as evidence that planets larger than $\sim 1.4-1.7\, R_\oplus$ have compositions more comparable to Neptune than the Earth, and are therefore not habitable in the traditional sense.

Very small planets will also have trouble maintaining even small and dense atmospheres against Earth-like insolations. \cite{Zahnle2017} find that a simple power law relationship between a body's escape velocity and its effective insolation ($I \propto v_\text{esc}^4$) can predict whether planets in the solar system (and some exoplanets) have atmospheres. According to this relation, a planet with the same insolation and density as the Earth would need to be larger than $0.8\,  R_\oplus$ to maintain a habitable atmosphere. However, we note that the Earth lies towards the inner edge of the LWHZ as calculated by \cite{Kopparapu2014}; it is possible that smaller planets could maintain Earth-like atmospheres further out.

\begin{figure}
    \centering
    \includegraphics[width=0.5\textwidth]{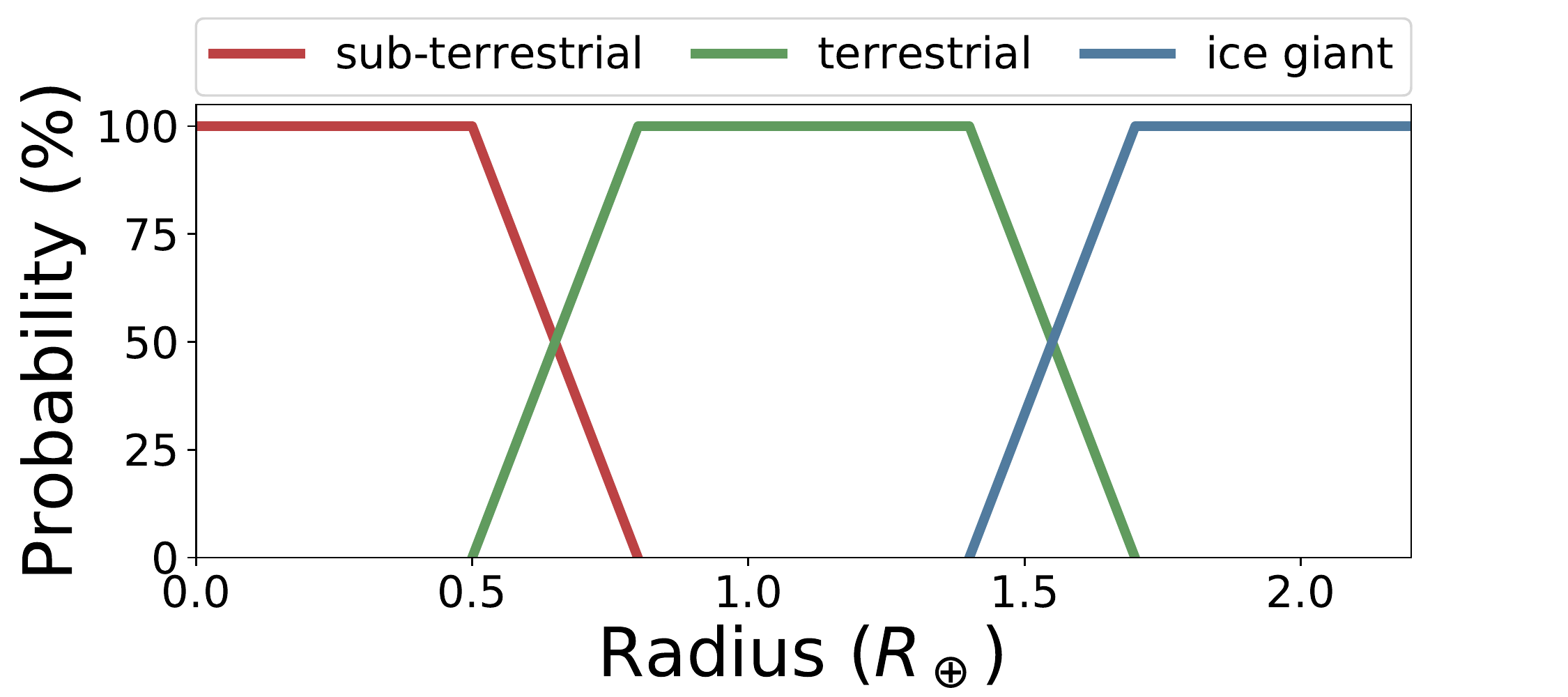}
    \caption{Our probabilistic scheme for classifying planets based on their radii. ``Sub-terrestrial'' planets are so small that they will lose their atmospheres to thermal escape under LWHZ levels of irradiation. ``Ice giants'' are so massive that they will form and maintain thick volatile envelopes. Only the ``terrestrial'' planets are neither too small nor too large to maintain a habitable atmosphere against irradiation.}
    \label{fig:classes}
\end{figure}

Taking both of these considerations into account, we assign one of three classes to each planet based on its radius: ``sub-terrestrial'' planets which are too small to be habitable, ``terrestrial'' planets which could have Earth-like atmospheres, and ``ice giant'' planets which are too large. There is likely some overlap between these categories; for example, planets slightly larger than $1.4\ R_\oplus$ or smaller than $0.8\ R_\oplus$ might still have an Earth-like atmosphere. To simulate this overlap we probabilistically assign each planet's class from its radius using the probabilities in Figure \ref{fig:classes}. The ``terrestrial'' class includes all planets with $0.8 < R < 1.4\ R_\oplus$ and a fraction of planets with $0.5 < R < 0.8\ R_\oplus$ or $1.4 < R < 1.7\ R_\oplus$.

\subsection{Mass}
To calculate each planet's mass, we rely on the empirical mass-radius relationships of \cite{Wolfgang2016}\footnote{Equation 2 and Table 1} (hereafter \citetalias{Wolfgang2016}), which are calculated for smaller ($< 1.6 \, R_\oplus$) and larger ($< 4 \, R_\oplus$) planets, reflecting the bimodal split in planet compositions. They do not treat mass as a deterministic function of radius, but rather model a \emph{distribution} of masses for each radius to capture the intrinsic variability in planet compositions.

We draw planet masses from truncated normal distributions defined by mean $\mu$ and variance $\sigma^2$, with minimum values of 0.01 $\mu$ and maximum values of $M_\text{pure Fe}$ - the mass of a pure iron composition as defined in \citetalias{Wolfgang2016}. The parameters of the distributions are:

\[
\mu,\sigma =
\begin{cases}
2.7 \, M_\oplus \, (R/R_\oplus)^{1.3} \ ,\ 1.9 \, M_\oplus     & (R \geq 1.6 \, R_\oplus) \\
1.4 \, M_\oplus \, (R/R_\oplus)^{2.3} \ ,\ 0.3 \mu             & (0.8 < R < 1.6 R_\oplus) \\
1.0 \, M_\oplus \, (R/R_\oplus)^{3.0} \ ,\ 0.3 \mu             & (R \leq 0.8 \, R_\oplus) \\
\end{cases}
\]

These are the values of $\mu$ and $\sigma$ fitted by \citetalias{Wolfgang2016}, with a few caveats:
\begin{enumerate}
    \item Their data did not allow the authors to determine $\sigma_M$ for the smaller planets; here, we arbitrarily choose $\sigma = 0.3 \mu$ (i.e., a 30\% spread in density).
    \item Only a few uncertain data points and upper limits were available for planets with $R < 0.8 R_\oplus$, so we instead assume approximately Earth-like densities.
    \item The large radius relationship was fitted for all planets with $R < 4 \, R_\oplus$, not just the ice giants; however, most of the precise data points nevertheless had $R > 1.6 \, R_\oplus$.
    \item We model a few planets as large as $10\, R_\oplus$, but this mass-radius relationship is likely not valid beyond $4\, R_\oplus$; indeed, it underestimates the mass of Jupiter ($\sim 11$ $R_\oplus$) by a factor of five. Ultimately, the overlap in masses and magnitudes between Earth-sized and Jupiter-sized planets is negligible when considering potential false positives for exo-Earth candidate detections.
\end{enumerate}

Finally, we expect that the empirical mass-radius relationship will be improved upon in coming years by the discovery of transiting rocky planets around low-mass stars by TESS \citep{Ricker2014}, precision radius measurements from CHEOPS \citep{Broeg2013}, and mass measurements through TTV or RV. By the time an exo-Earth direct imaging mission begins, observers should have a better understanding of the relationship between a planet's size, composition, and mass with which to interpret radial velocity or astrometric mass measurements.

\subsection{Eccentricity} \label{sec:eccentricity}
Planets in the solar system tend to have eccentricities smaller than 0.1, but multiple authors find evidence for a wider distribution of exoplanet eccentricities in both transit \citep{Kane2012} and radial velocity data \citep{Kipping2013}. \cite{Kipping2013} determine that the eccentricity distribution of several radial velocity detected exoplanets is well-described by a beta function, with $\alpha = 0.867$ and $\beta = 3.03$ - in which case $>50\%$ of planets have $e > 0.1$.

Eccentricity will have the effect of confusing the determination of a planet's orbit from a single epoch of imaging data, as a wider range of eccentric orbits could be consistent with the observed separation. Furthermore, planets which orbit near the inner or outer edge of the LWHZ may spend a fraction of their orbit outside of the zone. To determine which of these planets are EECs we use the mean flux approximation - assuming that a planet is habitable if the average insolation of its orbit is the same as a circular orbit in the habitable zone; \cite{Bolmont2016} find this to be an adequate approximation for planets with low or modest eccentricities receiving a mean flux equal to the Earth's, while highly eccentric planets tend to freeze out. However, it is possible that this approximation is not valid for eccentric planets receiving a lower mean flux.

We use the beta distribution of \cite{Kipping2013} to draw planet eccentricities and compute the projected separation accordingly, though we truncate the distribution beyond $e > 0.8$ (0.5\% of planets) due to the additional computing time required to solve Kepler's equation for highly eccentric orbits.

We note that multi-planet systems such as our own tend to have more circularized orbits \citep{VanEylen2015}. In principle, if a planet is detected with a companion then it is \emph{a priori} less likely to have an eccentric orbit, and it should be easier to determine whether the planet is an EEC. In the scope of this paper, however, we treat all planets as the only member of their system.

\subsection{Albedo} \label{sec:types}

We consider two different prescriptions for simulating the geometric albedo: in Cases 1-5 we assume a monochromatic albedo drawn from a broad uniform distribution, while in Cases 6 and 7 we assume a spectral albedo model which depends on the planet's class and position with respect to the LWHZ.

Constraining the actual distribution of planet surface and atmospheric properties is one of the goals of future imaging missions, so it might seem backwards to interpret these observations by assuming the underlying distribution of geometric albedos as a prior. Nevertheless, such an assumption is necessary in order to infer a planet's size from photometric data.

A uniform prior represents a conservative approach to the problem. We note that most planets in the solar system have geometric albedos ranging from 0.2 to 0.7 across the UV to NIR wavelength range, with the exception of Mercury's very low albedo. Therefore for most cases we draw a monochromatic geometric albedo for each planet uniformly from 0.2 to 0.7.

However, the proposed designs of direct imaging missions allow for the simultaneous observations of a planet in 2-3 photometric bands, in which case color information would be available with no additional overhead. An efficient characterization strategy would make use of this color information to discriminate between EECs and their false positives, but to do so the observer must assume some prior knowledge about the diversity of planetary atmospheres and surfaces. As an example, we assume that all planets approximately reflect one of four solar system analogs - Earth, Venus, Mars, or Neptune - based on their radius-based classification and orbit as described in Table \ref{tab:types}.

We accessed model spectra through the Virtual Planetary Laboratory\footnote{http://depts.washington.edu/naivpl/content/vpl-spectral-explorer} for the Earth \citep[][scaled from quadrature]{Robinson2011}, Venus dayside, and Mars (no publications listed). For Neptune we use the planet's observed geometric albedo as provided by \cite{Madden2018} from 450-2500 nm, and set $A_g = 0.6$ from 300-450 nm \citep[e.g.,][]{Mallama2017}.

\begin{figure}
    \centering
    \includegraphics[width=0.5\textwidth]{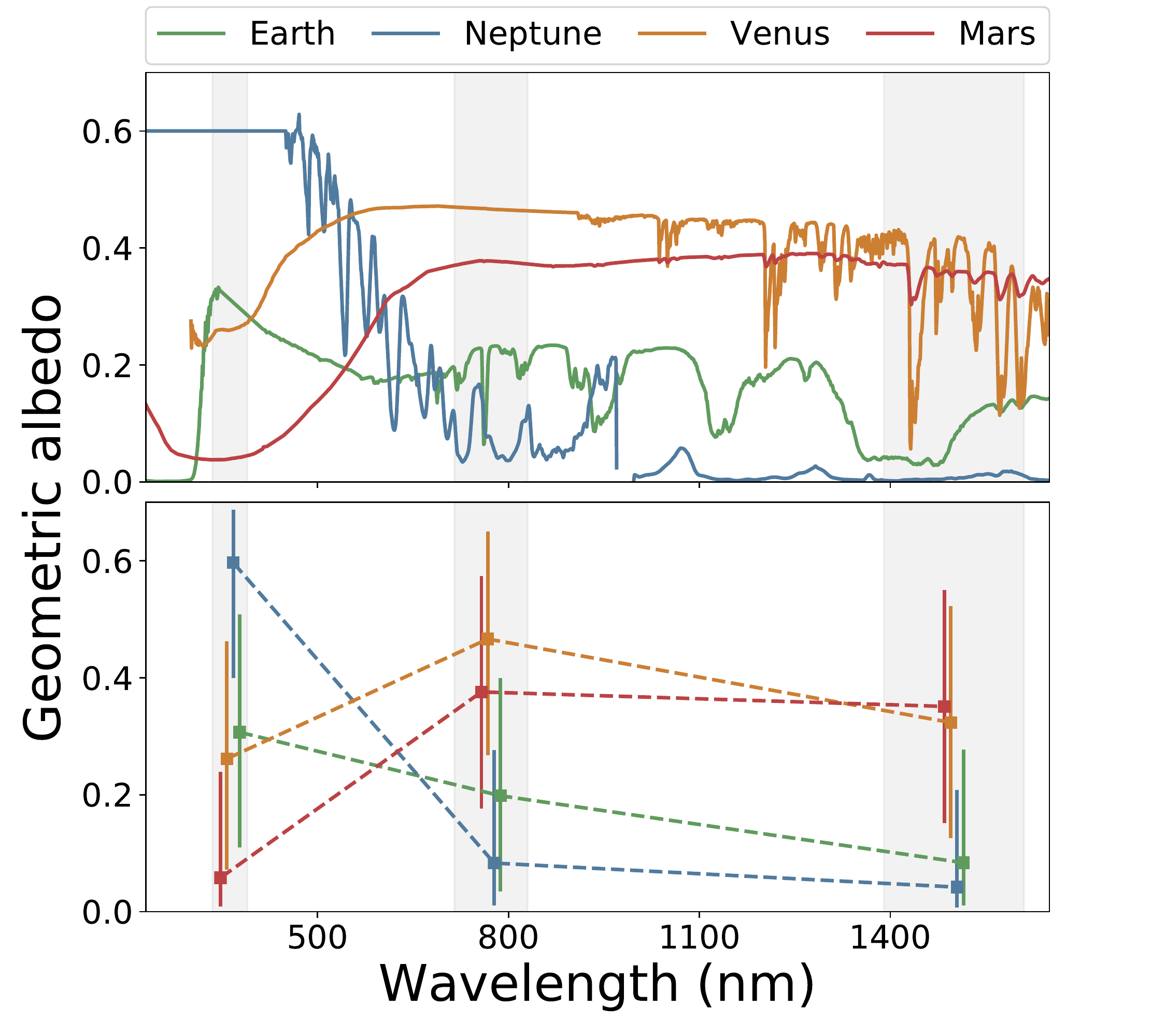}
    \caption{(Top) The set of base models which we use to simulate geometric albedos under Cases 6 and 7 only, according to the scheme in Table \ref{tab:types}. (Bottom) The actual range of simulated spectra for each type of planet as observed in three photometric bandpass (with wavelength offsets for visibility). We apply moderate differences to each planet's spectrum to simulate the underlying physical diversity, but broad differences between the four groups can still be seen.}
    \label{fig:spectra}
\end{figure}

To simulate both model uncertainty and physical diversity among these solar system analogs, we allow the spectral albedos to vary by $\pm 0.1$ monochromatically and $\pm 0.05$ in each photometric bandpass. Additionally, we enforce a lower limit of 0.001 - to ensure that none of our planets are perfect blackbodies - and an upper limit of 0.7, slightly more than the upper limit for the geometric albedo of a Lambertian sphere. The range of spectral models for each category of planet is demonstrated in Figure \ref{fig:spectra}.

Finally, we consider the bandpasses in which the planets are observed. For reference, LUVOIR's proposed coronagraphic instrument would be able to observe simultaneously in $10\%$ bandpasses of each of its three channels \citepalias{LUVOIR2019}. We choose wavelength ranges near the centers of each channel: 335-390 nm, 715-830 nm, and 1390-1610 nm.

It is likely that a wide variety of terrestrial planets exist; indeed, all four of the terrestrial planets in the solar system are distinct in surface reflectance and atmospheric absorption. Both of the ice giants, however, have similar albedo distributions. This suggests that some understanding of the albedo distribution of ice giant analogues can be attained during the coming decade. Modeling efforts to understand the composition and appearance of sub-Neptune type planets are already underway \citep[e.g.,][]{Hu2015}. New observatories such as JWST, WFIRST, and ELTs could provide observational tests of these models - the first through eclipse and transit spectroscopy of warm and hot Neptunes, and the latter two through direct imaging of ice giants at wide separations.

\begin{deluxetable}{lll}
\tablecaption{A list of the solar system analogs which we simulate each planet's spectrum under Cases 6 and 7. The assumed model for the planet's spectrum depends on its class/size and location with respect to the liquid water habitable zone, as defined by the runaway and maximum greenhouse limits. \label{tab:types}}
\tablehead{Class & Location & Assumed model}
\startdata
terrestrial & in LWHZ & Earth \\
sub-terrestrial & anywhere & Mars  \\
terrestrial & exterior to LWHZ & Mars \\
terrestrial & interior to LWHZ & Venus \\
ice giant & anywhere & Neptune \\
\enddata
\end{deluxetable}

\section{Mock surveys: methodology} \label{sec:mock_survey}

The LUVOIR team has released a final report \citepalias{LUVOIR2019} which include estimates for the number of planets which LUVOIR could detect as a function of planet radius, insolation, and host spectral type. In this section, we produce a candidate sample based on these estimates, characterize each planet therein, and report on the efficiency of each strategy outlined in Table \ref{tab:cases} for properly identifying exo-Earth candidates. While we rely on the yield estimates and stellar targets of \citetalias{LUVOIR2019} for our mock survey, we use these only as baseline estimates for the yield of a generic, hypothetical direct imaging survey; we do not attempt to reproduce the results of the report or to assess the mission design.

\subsection{Generating the candidate list}

To properly generate a list of planets which could be detected by a coronagraphic imaging mission requires a careful treatment of the instrument design, survey strategy, and additional sources of noise (e.g., dust) which is outside of the scope of this work. \citetalias{LUVOIR2019} have performed such an analysis (following the methodology laid out by \cite{Stark2014}) so we use their results to generate our candidate list\footnote{see Figures 3.1 \& 3.13 and Tables 8.7 for details on the target list, yield estimates, and instrument parameters}.

We acquired one of the simulated samples of host stars upon which the LUVOIR-A yield estimates are based, including masses, luminosities, and distances for 287 stars (C. Stark, private correspondence). To assign a radius to each star we use a simple scaling relation \citep{Hansen2004}:
\[
(R_*/R_\odot) = (M_*/M_\odot)^{0.8}
\]

While this simulated list will not be the final target list of LUVOIR-A, it generally represents the diversity of host star properties which such a survey would encounter. Next, we use the method outlined in Section \ref{sec:prior_sample} to generate a large sample of planets around these stars, and draw from that sample at random until the yield estimates for LUVOIR-A have been satisfied for each bin in spectral type, planet size, and insolation.

To ensure that the planets we simulate are detectable, we also enforce the same separation and brightness limits as \citetalias{LUVOIR2019}. Namely, we only include targets which are brighter than a planet-to-star contrast ratio of $2.5\times10^{-11}$ and which are detected between 24--440 mas - approximately the working angles of the proposed LUVOIR-A coronagraph at 500 nm.

The yield estimates of the LUVOIR-A architecture project the discovery of $\sim 450$ planets, $\sim 50$ of which would be exo-Earth candidates. For our purposes, the actual number of planets is not relevant - only their relative abundance by size, distance, host spectral type, etc. - so we improve the accuracy of our results by inflating the yield estimates unilaterally by a factor of fifty. 

\subsection{Survey cases}

We run our mock surveys under seven cases representing the different data which could available to the observer. A brief description of each can be found in Table \ref{tab:cases}.

\subsubsection{Case 1: Detection data only}
Under Case 1, the planet's existence is entirely unknown before its direct imaging detection, and the observer only measures its apparent separation and monochromatic magnitude. Following \citetalias{LUVOIR2019}, we assume a signal-to-noise ratio of 7. If photon noise is dominant, the uncertainty on the planet's centroid position is described by:

\[
\sigma_c = \frac{\text{FWHM}}{\text{SNR}}
\]

Assuming an effective wavelength of 500 nm, then $\sigma_c$ = 3.8, 1.9, and 1.0 mas for the 4-meter HabEx and 8- or 15-meter LUVOIR architectures. This uncertainty will be further affected by the pointing stability of the telescope during observations and the wavelengths at which the planet is observed; to be conservative, we choose $\sigma_c$ = 3.8 mas. Finally, we simulate the planet's detection by re-drawing its magnitude and apparent separation from normal distributions with widths defined by these uncertainties.

\subsubsection{Case 2: Additional radial velocity detection}
It has been emphasized that a radial velocity search for nearby Earth twins would be an important precursor to a space-based direct imaging mission \citep{Dressing2019}, but the detection of Earth analogs around Solar twins is beyond the reach of current instrumentation. Doing so would require a significant investment of time on major observing resources and new methods to correct for systematic sources of noise such as stellar jitter \citep{Plavchan2015}. 

To investigate the potential benefits of a precursor radial velocity search for interpreting planet detections, in Case 2 we simulate the direct imaging detection of a planet (Case 1) along with a measurement of its orbital parameters and radial velocity semi-amplitude. We set a baseline uncertainty of 5 cm/s on the measurement of the radial velocity semi-amplitude $K$ - this value is chosen to be smaller than the value for an Earth twin ($\sim 10$ cm/s), but not negligible in comparison. Additionally we allow a conservative 10\% uncertainty on the measurement of the orbital period, and a loose $\pm 30^\circ$ constraint on the planet's mean anomaly.

Several planets - including small planets in the LWHZ which could masquerade as exo-Earths candidates - will have $K < 5$ cm/s. In these cases we assume the planet is undetected, so no constraints on its orbit are available. An upper limit of $10$ cm/s is enforced so that the non-detection also carries useful information about the planet's size.

\subsubsection{Case 3: Constraining the orbital plane using debris disks measurements}
It is expected that several nearby systems contain exozodiacal dust disks near their habitable zones \citep{Ertel2018}. This dust may be a significant source of background and confusion noise for future direct imaging surveys and could be a key driver of aperture size, with larger apertures collecting less background light within their resolution element \citep{Roberge2012}. However, these disks could also carry useful information about the orientation of the system which could help to establish a newly detected planet's orbit, and would require no additional resources to observe.

Observations of larger protoplanetary disks have yielded tight (better than $1^\circ$) constraints on the disk inclination and orientation \citep[e.g., HL Tau,][]{Alma2015}. While the orientation of exozodiacal disks may be harder to constrain if they are faint or exceed the outer working angle of the coronagraph - and while not all systems may have substantial disks at all - we here consider the ``optimistic'' case where every system has a disk, and the orientation parameters ($\cos(i)$ and $\Omega$) can be constrained to $\pm 1\%$.

In principle, if the orientation of the debris disk can be tightly constrained and the planet shares \emph{exactly} its orbital plane with a circular orbit, then a single precise measurement of the planet's apparent separation could be sufficient to determine its phase and semi-major axis. However, the orbital inclinations of solar system planets deviate by up to $5^\circ$ from the Solar spin axis - and larger misalignments could be possible in other systems - while several of our simulated planets have eccentric orbits.

We can still infer some information about the planet's orbit given the orientation of the disk and some prior knowledge about how misaligned planetary systems tend to be. In Case 3, we use the orientation of a contemporaneously detected disk as a prior constraint on the mean orbital plane of the system, but allow for a difference of $\pm 0.2$ between $\cos(i)$ and the longitude of the ascending node ($\Omega$) of the two components. We then interpret the observations using the same data as in Case 1, but treating the two dimensions of the separation vector separately.

\subsubsection{Cases 4 and 5: Multiple revisits and astrometric mass measurements}
Constraining the planet's orbital parameters will require multiple revisits spaced over the orbital period, so revisits will likely be folded into the observing strategy of future direct imaging missions. However, even within the habitable zone there will be numerous potential false positives for EECs, a fact which may limit the practical benefit of revisiting every system.

\cite{Guimond2019} have shown that $\sim 3$ revisits will be sufficient to constrain the orbital parameters with better than $10\%$ precision. In Case 4 we assume that the system has been revisited enough times for the planet's orbital parameters to be constrained with comparable precision. Specifically, we assume that the semi-major axis is measured to $\pm 10\%$, the eccentricity to $\pm 0.05$, and the orbital phase to $\pm 15^\circ$.

An ancillary benefit of revisiting targets would be the ability to measure the star's astrometric motion about the system's center of mass. Measuring the astrometric semi-amplitude $\theta$ would allow observers to determine the planet's mass and potentially to identify which planets are too small or large to be habitable. In Case 5 we assume that, in addition to the constraints provided in Case 4, the observer can measure $\theta$ with a precision of $\pm 0.1$ $\mu$as. This is the targeted astrometric precision of the High Definition Imager (HDI) instrument with the LUVOIR-A aperture \citepalias{LUVOIR2019}, and approximately $1/2$ of the amplitude induced by an Earth twin at a distance of 15 parsecs. If $\theta < 0.1$ $\mu$as, an upper limit of $0.2$ $\mu$as is enforced instead.

\subsubsection{Case 6: Color measurements}

Future direct imaging missions could be able to observe simultaneously in multiple bandpasses, in which case measurements of the planets' colors would be available as soon as they are detected. Color information could be used for preliminary planet classification; for example, \cite{Batalha2018} predict that it will be possible to differentiate between cloudy and cloud-free Jovian planets with reflected light imaging in three filters using WFIRST or ELTs.

However, color information is only useful for inferring the planet's properties if we make a prior assumption about the spectral albedos of extrasolar planets. Some observational constraints are currently available through eclipse observations of close-in giant planets, but for smaller planets we can only rely on planets in the solar system and theoretical models of the surfaces and atmospheres of known transiting and RV-detected planets. It is therefore worthwhile to determine what effect an approximate prior understanding of spectral albedos would have on the interpretation of direct imaging data.

In Case 6 we simulate planets with spectral albedos reflective of (though not identical to) solar system analogues; our detailed assumptions are described in Section \ref{sec:types}. Each planet is observed in a 10\% bandpass at the center of three wavelength channels, with a signal-to-noise ratio weighted by the square root of the bandpass-integrated flux (i.e., photon noise). The signal to noise integrated across all three bandpasses is 7, as in Case 1. If S/N < 2 in a given bandpass (typically in the UV or infrared), then a 2$\sigma$ upper limit is enforced instead.

\subsubsection{Case 7: Maximum information}
In the final case we consider all of the information which a larger telescope would be able to acquire on a target after several revisits. These include measurements of the orbital parameters and astrometric semi-amplitude as well as brightness measurements in three bands. We assume the same measurement precision and sensitivity limits as in Cases 4 and 5, and we interpret the color information following the method of Case 6.

\subsection{Classifying the targets}
\begin{figure*}
    \centering
    \figuretitle{\large How confidently can we identify exo-Earth candidates in mock observations?}
    \includegraphics[width=0.65\textwidth]{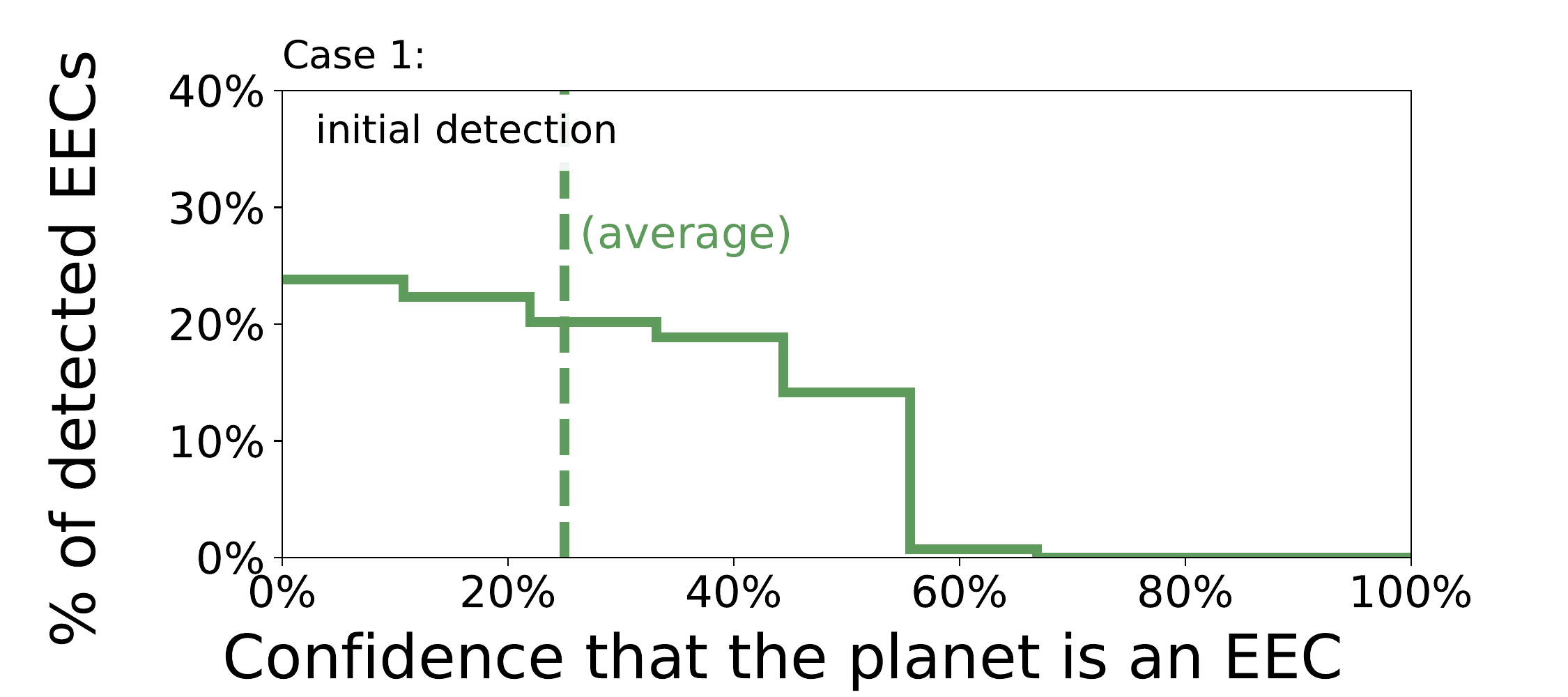}
    \includegraphics[width=\textwidth]{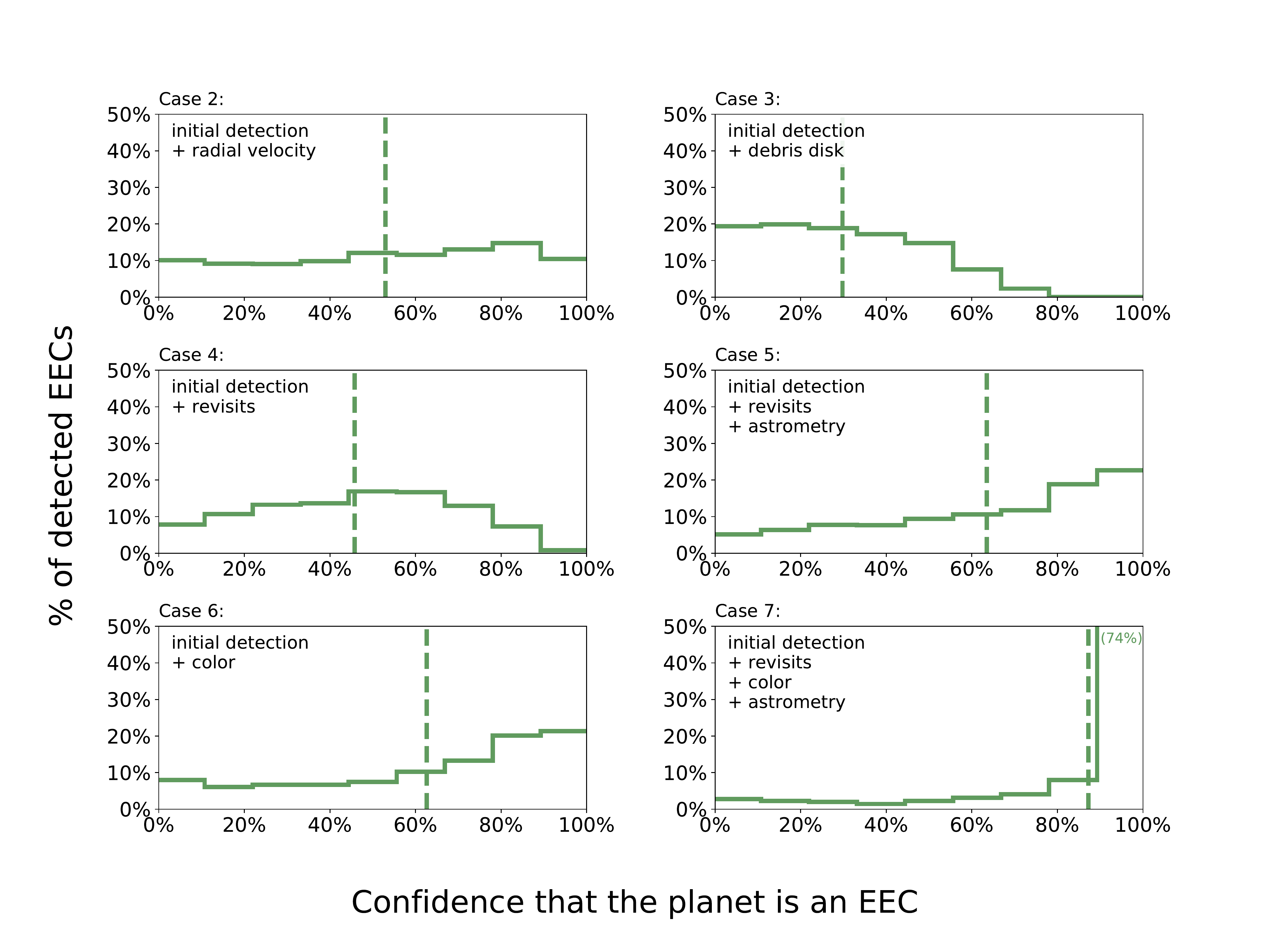}
    \caption{We simulate the detection of $\sim 2500$ EECs around nearby stars, then infer their properties from the mock data under each of the cases in Table \ref{tab:cases}. Above we plot the probability - as inferred by an uninformed observer - that each detected planet is an EEC, as well as the average value (dashed line). In the ideal case, this value would be 100\% for all EEC targets, but typically it is smaller because of the limited data available to the observer. With additional data (Cases 2--7), the observer can be more confident that the detected planet is an EEC.}
    \label{fig:P_EEC}
\end{figure*}

Once we have constructed a sample of targets and simulated their detection, we classify them using the inference framework described in Section \ref{sec:framework}. This yields for each planet the likelihood, according to the observer, that the planet is an EEC or a false positive. In Figure \ref{fig:P_EEC} we plot the likelihood that the planet is an EEC for each simulated EEC in the sample. Since the observer does not know \emph{a priori} that these planets are EECs, we see that they cannot make a confident identification upon the planets' initial detections, but given additional data or multiple revisits they can identify several EECs with confidence.

\begin{figure*}
    \centering
    \figuretitle{\large What kinds of planets could be mistaken for exo-Earth candidates?}
    \includegraphics[width=0.65\textwidth]{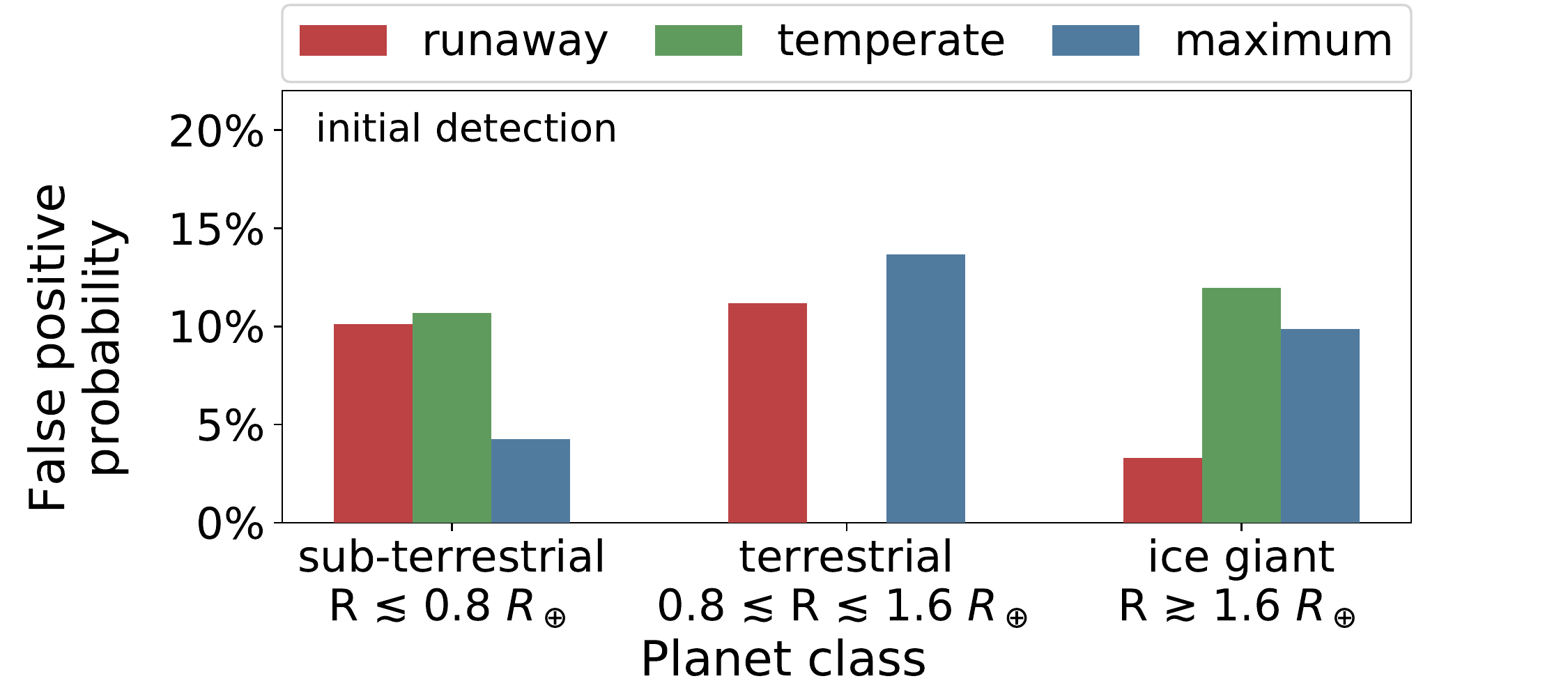}
    \includegraphics[width=\textwidth]{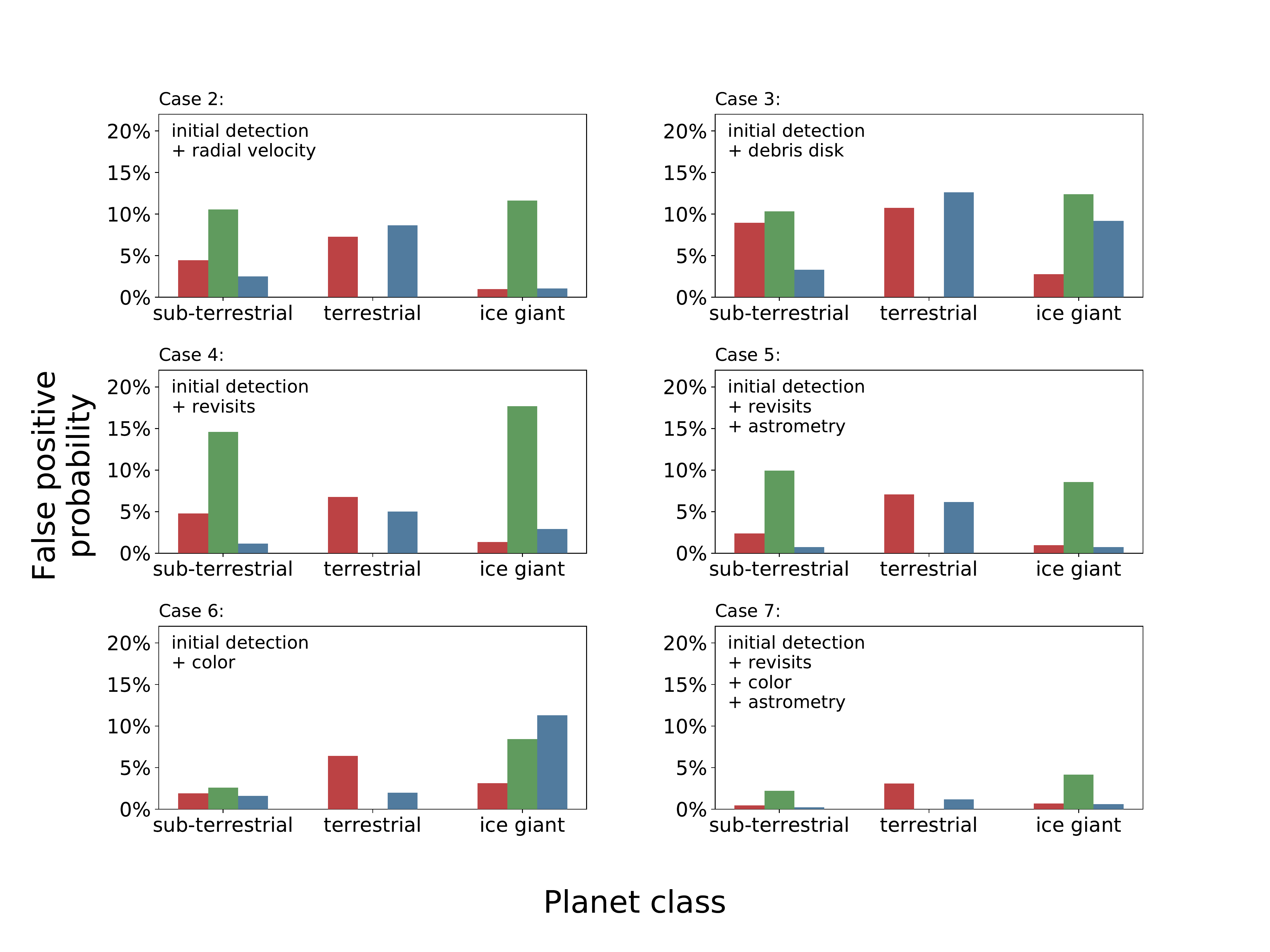}
    \caption{We simulate the detection of $\sim 2500$ EECs around nearby stars, then infer their properties from the mock data under each of the cases in Table \ref{tab:cases}. Above we plot the sample-averaged probability - as inferred by an uninformed observer - that the planet is instead a false positive with a non-habitable class or orbit. Additional data will suppress the false positive probability - for example, fitting the orbit with finite precision will reduce the likelihood that the planet is outside of the habitable zone.}
    \label{fig:classification}
\end{figure*}

We can break down the false positive probability by size and orbit, determining for each observed EEC the inferred likelihood that it is, for example, a sub-Neptune on an orbit exterior to the habitable zone. In Figure \ref{fig:classification} we plot this probability for each combination of class and orbit averaged over the sample of observed EECs. This plot illuminates the key sources of ambiguity in classifying EECs. For example, we see that it is difficult to distinguish between an EEC and a sub-terrestrial planet with a temperate orbit, or a planet which has the proper size but lies just interior or exterior to the habitable zone.

\section{Mock surveys: results} \label{sec:results}

\subsection{Which planets are ``false positives''?} \label{sec:false_positives}

Figure \ref{fig:classification} demonstrates that true EECs share the observable parameter space with a wide range of planets both within and outside of the LWHZ. These can be broadly separated into: (i) planets which are too small or large to be habitable, but are yet in the habitable zone, (ii) planets which are of the proper size to be habitable, but are not in the habitable zone and (iii) planets which are both of the wrong size and not in the habitable zone. Each of these categories are approximately equal in their potential to masquerade as EECs.

As we demonstrate in Figure \ref{fig:P_EEC}, the observer will typically be unable to distinguish between true EECs and their many potential false positives given just the data necessary for the planet's detection. Even when a planet which is \emph{in fact} an EEC is detected, the observer will only be able to make this determination with $<50\%$ confidence.

\subsection{Do constraints on the orbit help to identify EECs?}

Since most of the ``false positives'' are planets outside of the habitable zone, it stands to reason that measurements which constrain the planet's orbit would be useful for identifying EECs. Indeed, if the planet can be independently detected through RV, and its period constrained with $10\%$ precision, then it can typically be constrained to the habitable zone with $>80\%$ confidence. However, approximately 25\% of EECs remain below our 5 cm/s detection limit, in which case the orbit cannot be established.

Observing the planet multiple times over an orbital period will help to rule out planets outside of the habitable zone with similar confidence, assuming 10\% uncertainties on the orbital parameters. Nevertheless, even if a planet can be constrained to the habitable zone, it may yet be too large or small to be habitable. In general, constraints on the orbit and phase will only allow the observer to distinguish EECs from temperate sub-terrestrial planets or ice giants with $\sim 50\%$ confidence.

\subsection{Do constraints on the mass help to identify EECs?}

We find that measurements of the astrometric semi-amplitude $\theta$, when combined with magnitude measurements, can modestly increase the observer's ability to identify EECs. For example, our average confidence for identifying EECs given multiple revisits to determine the orbit (Case 4) improves by about 18\% if we include an astrometric measurement or upper limit on the planet's mass (Case 5), and several individual EECs can be identified with very high confidence. These could be the highest priority targets for deeper spectroscopic follow-up.

Under Case 2, if both the measured period and radial velocity semi-amplitude $K$ are used to constrain the planet's properties then EECs can be identified with an average confidence of 52\%. If the measured value of $K$ is ignored, however, then this confidence drops to $40\%$. In other words, the measurement of the planet's mass affords an extra 12\% confidence that the planet is an EEC.

Constraining the orbit generally reduces the potential for false positives more than constraining the mass, in part because constraints on the planet's size are available based on its brightness alone. On the other hand, the relationship between radius, mass, and composition is more well-understood than the prior distribution of planet albedos (which almost certainly is \emph{not} uniform). An inference about the planet's composition made from the its mass could be more reliable than one made from its apparent magnitude.

\begin{figure*}[]
    \centering
    \figuretitle{\large Does color information help to identify exo-Earth candidates?}
    \includegraphics[width=\textwidth]{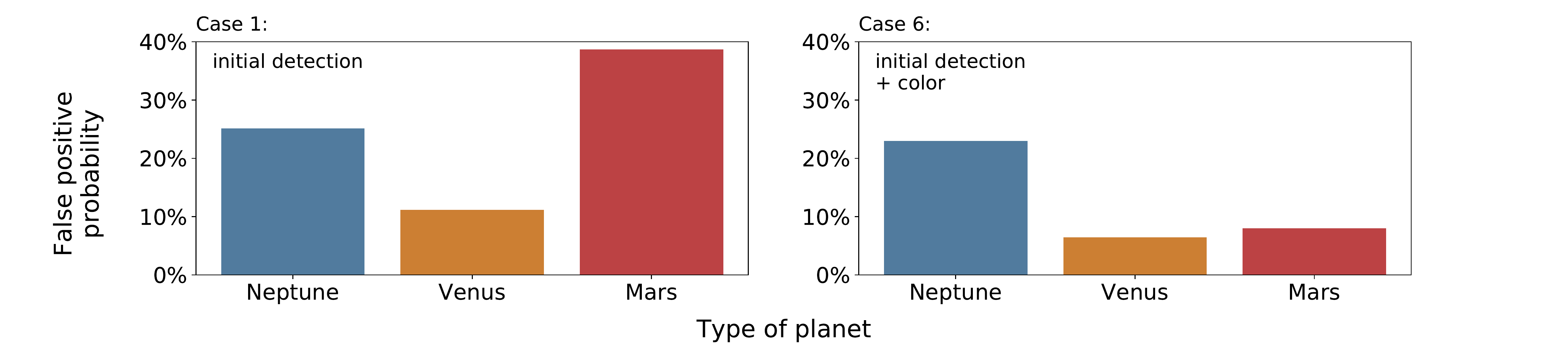}
    \caption{We test whether low-to-moderate S/N color measurements would help to identify EECs, assuming that all planets look \emph{approximately} like a solar system analog depending on their class and orbit (Table \ref{tab:types} and Figure \ref{fig:spectra}). Plotted is the typical inferred probability that the EEC is actually a false positive with a Neptune, Venus, or Mars-type surface and atmosphere. Without color information (left) it is difficult to distinguish Earth analogs from smaller or cooler Mars-like planets, or larger Neptune-like planets. With color information (right), the slope of the optical spectrum provides a useful discriminant between the Earth and Mars, but does little to reduce the ambiguity due to larger Neptunes.}
    \label{fig:types}
\end{figure*}

\subsection{Do color measurements help to identify EECs?}

\begin{figure}
    \centering
    \includegraphics[width=0.5\textwidth]{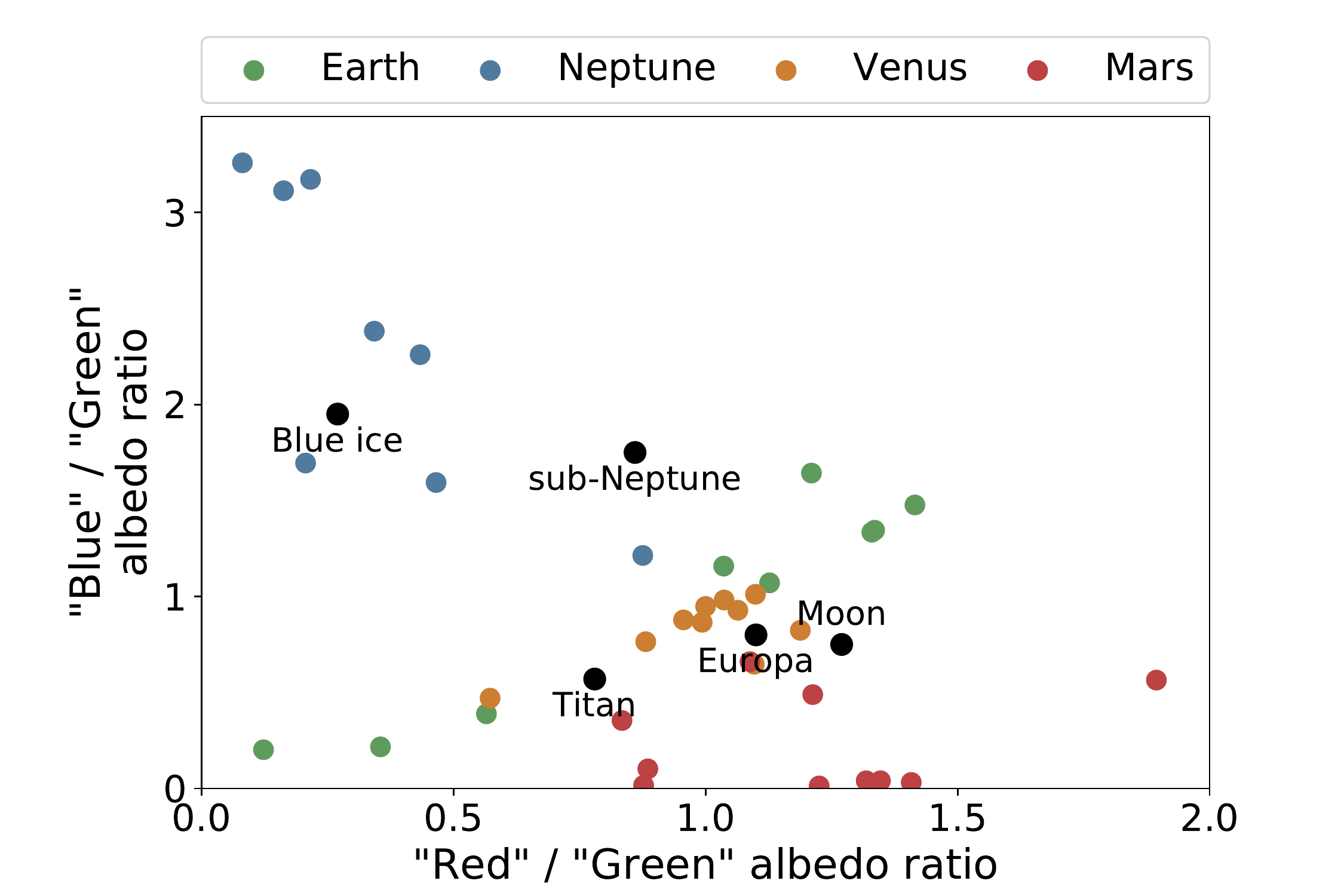}
    \caption{We plot a set of simulated Earth, Neptune, Venus, and Mars analogs in color-color space - each planet's spectrum is modified from the solar system model to simulate physical diversity. We also plot data points from \cite{Krissansen-Totton2016} (Figure 2) who calculate these color ratios for a large diversity of solar system bodies, materials, and exoplanet models. While we consider fewer base models than these authors, our simulated planets cover a similar range in color space.}
    \label{fig:colors}
\end{figure}

In Figure \ref{fig:types} we plot the average inferred probability that an EEC is similar to the Earth, Neptune, Venus, or Mars for Cases 1 (no color) and 6 (color). We find that color information is useful for distinguishing between spectra with positive versus negative slopes between the UV and visible channels. Specifically, adding a color measurement allows the observer to distinguish between Earth and Martian analogs particularly well - if small planets in the habitable zone tend to look like Mars, color information will be a valuable discriminant. According to Figure \ref{fig:P_EEC}, observing the color \emph{and} constraining the planet's orbit could allow the observer to identify most EECs with $> 80\%$ confidence.

We choose solar system planets as our templates as they cover a relatively broad range of insolations and planet sizes, and with the exception of very hot exoplanets they remain the only planets for which albedo measurements are presently available. In our set of models shown in Figure \ref{fig:spectra}, the Earth stands out due to its modest scattering slope in the optical versus a much stronger feature in Neptune's spectrum, or opposite features for Venus and Mars. Yet a habitable planet does not need to look like the Earth, and indeed there is evidence that during the Archean the Earth had a substantially redder appearance due to organic haze particles \citep[e.g.,][]{Arney2016,Arney2017}. Similarly, the solar system provides no examples of an Earth-sized planet beyond the habitable zone. Here we assume such worlds have a Martian appearance, but this is likely inaccurate for icy worlds or planets with dense atmospheres.

We compare our range of simulated spectra to the models of \cite{Krissansen-Totton2016}, who compute optimal photometric bandpasses to distinguish between several different examples of potential exoplanet reflectance spectra. Their optimized bandpasses are 431-531 nm ("blue"), 569-693 nm ("green"), and 770-894 nm ("red") - different than those used in this work. In Figure \ref{fig:colors} we place our models on a color-color plot similar to Figure 2 in the cited work, along with a subset of the models considered therein. We find that our solar system analogs with simulated physical diversity cover a comparable range in color space, so we believe that we adequately represent a diversity of planet appearances even though our range of base models is limited.

We stress that this result is sensitive both to our prior assumptions and to the bandpasses in which we choose to observe our targets. More work must be done to understand the potential diversity of terrestrial planets and to determine which photometric bandpasses are optimal for distinguishing them from EECs \citep[e.g.,][]{Krissansen-Totton2016}. In Section \ref{sec:future_priors} we discuss how observational constraints on the albedos of potential false positives could be derived within the coming decade. These new discoveries can then be folded into our Bayesian framework, and the results of Case 6 suggest that doing so could allow for the confident distinction between false positives and true EECs on the basis of color information.

\subsection{Can a debris disk be used to constrain the orbital plane?}

We find that measuring the orientation of the debris disk provides relatively little information about the planet's orbit. Specifically, we are on average only $\sim 5\%$ more confident that the observed EECs lie within the habitable zone when we have measured the orientation of the disk. The benefits are slightly greater in cases where the system is observed from a ``pole-on'' orientation ($|\cos (i)| > 0.5$), in which case the uncertainty in the centroid measurement translates to a smaller uncertainty in the semi-major axis versus the ``edge-on'' cases ($|\cos(i)| < 0.5$). Nevertheless, unless the centroid precision is much better than $3.8$ mas and the observed systems are at least as well-aligned as the planets in the solar system, measuring the disk orientation will generally not allow an observer to constrain the orbit without revisits.

\subsection{Can EECs be identified given maximal photometric information?}
In Case 7 we assume that the observer has revisited the system multiple times to constrain the orbit with three-band photometry and has additionally measured (or placed upper limits on) the astrometric motion of the star due to the planet. This is the most information which could be obtained for the typical system without substantial follow-up or precursor observations, and we find that it would allow the observer to confidently identify most EECs, with an average confidence of 87\%. This suggests a promising roadmap toward selecting targets for follow-up, but as in Case 6 this result is dependent on the observer's prior assumptions about exoplanet reflectance spectra.

\begin{figure*}[ht]
    \centering
    \figuretitle{\large \large Efficiently surveying planets for spectroscopic H$_2$O absorption}
    \includegraphics[width=\textwidth]{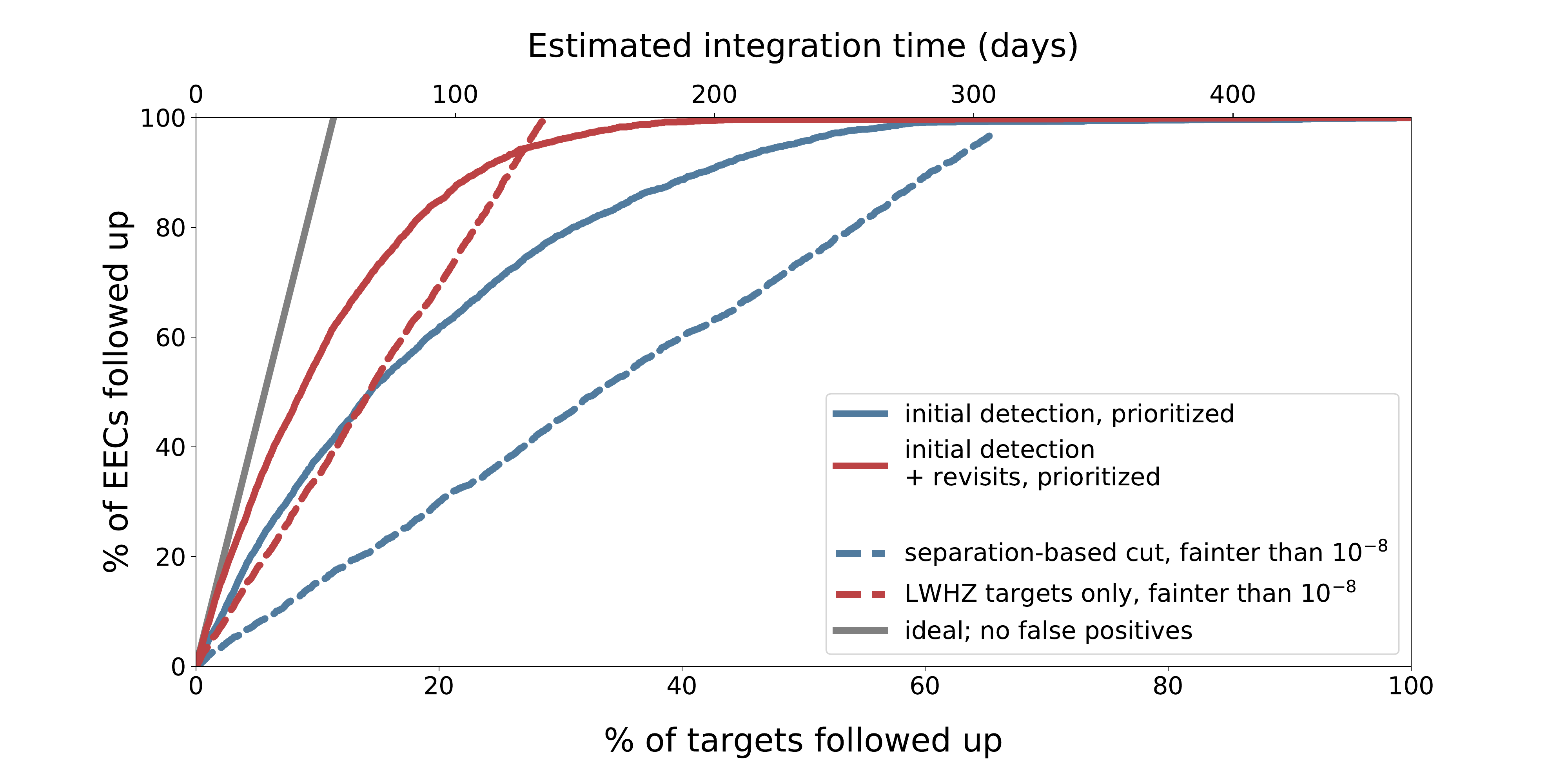}
    \caption{The efficiency of different post-detection follow-up strategies (e.g., to search for water absorption), quantified as the percentage of all targets which must be followed up before a given percentage of EECs have been covered; the ordinate is proportional to the amount of time required for follow-up. The gray line represents ideal survey efficiency, where no false positives are re-observed. The colored lines represent strategies which make use of the detection data only (blue) or multiple revisits to establish the orbit (red). The dashed lines are non-prioritized (blind) approaches which first remove very bright or widely-separated planets, or planets whose orbits are constrained to be outside of the LWHZ. The solid lines are prioritized approaches in which we observe the planets which are most likely to be EECs first. The upper axis estimates the amount of time required to search each planet's atmosphere for water absorption, assuming 25 hours of integration time per target.}
    \label{fig:efficiency}
\end{figure*}

\subsection{Would Bayesian prioritization improve follow-up efficiency?} \label{sec:efficiency}

A practical way to frame our results is in terms of follow-up efficiency. A logical next step after detecting a directly imaged planet (and optionally constraining its orbit) would be to search for water absorption in a narrow part of the spectrum to further test its habitability. However, unless a strategy is employed for prioritizing or pruning the target list, a significant amount of time will be spent ``following-up'' non-habitable planets. Indeed, \cite{Kawashima2019} have shown that 3-10 hours may be required to confidently detect H$_2$O absorption in an exo-Earth atmosphere from 5 parsecs using a 10-meter telescope. For the typical target observed by the 15-meter LUVOIR-A from 15 parsecs, approximately $\sim 25$ hours may be required.

Our inference framework allows us to prioritize targets based on the likelihood that they are, in fact, exo-Earth candidates. After probabilistically classifying each observed target - including both EECs and non-EECs - we prioritize them by the probability that they are true EECs and submit them for follow-up observations with the most likely candidates first. In Figure \ref{fig:efficiency} we summarize the efficiency of this prioritization strategy for Cases 1 (initial detection only) and 4 (multiple revisits).

We compare our Bayesian approach to two non-prioritized follow-up strategies: first, removing all targets whose projected separations are wider than the maximum greenhouse limit, and second, removing all targets that are found to be outside of the habitable zone after multiple revisits. In both of these cases we also remove targets with a contrast ratio brighter than $10^{-8}$. These separation- and magnitude-based cuts exclude $\lesssim 6\%$ of bright or eccentric EECs but several larger or cooler false positives. The remaining targets are then followed-up blindly. Finally, we plot a line representing a perfectly efficient follow-up strategy with no false positives.

We find that prioritizing the targets using our Bayesian framework allows us to re-observe them with a much greater efficiency - using the \emph{same} data - than a blind approach. In fact, a prioritized approach using just the detection data is initially \emph{as efficient} as taking a blind approach after each planet's orbit has been characterized. We can discuss these efficiency gains in terms of integration time. Assuming 450 targets have been discovered in the initial census and 25 hours of integration time are required to search each planet for water, the results in Figure \ref{fig:efficiency} suggest that probabilistic target prioritization could reduce the required amount of integration time to follow-up 50\% of EECs by 28 days (if the orbits have been precisely constrained) to 95 days (if only the detection data are available). Since this prioritization scheme does not require additional data on the system, it could be naturally folded into the survey strategy of any direct imaging mission.

\subsection{Is the Bayesian approach always appropriate?}
In principle, a Bayesian prioritization scheme should always be superior to a blind follow-up strategy because it leverages additional information about exoplanet demographics. In practice, this prior knowledge is always incomplete and potentially inaccurate, which could make Bayesian prioritization risky in the sense that it might fall prey to unexpected types of false positives. For example, to produce planets for our mock surveys we use the yield estimates of \citetalias{LUVOIR2019} which include more sophisticated treatments of the mission design and a different estimation and extrapolation of \emph{Kepler} occurrence rates - but then we use our own algorithm to probabilistically classify each planet. Therefore, the prior assumptions we make when characterizing these simulated planets are not exactly representative of their true parent distribution, which makes our prioritization less efficient.

Indeed, we see in Figure \ref{fig:efficiency} that if the planets' orbits are known, then the blind approach to prioritizing targets is \emph{more} efficient than the Bayesian prioritized approach for the last $\sim 5\%$ of EECs. However we argue that this is realistic - the observer's prior assumptions will always differ from reality to a degree - and we note that our approach is still efficient despite this mismatch. For this reason we believe Bayesian prioritization will yield \emph{overall} better results, but observers might prefer a blind approach for low priority targets.

\subsection{What priors can be improved in the coming decades?} \label{sec:future_priors}
The accuracy of our probabilistic classification depends on the extent and accuracy of existing knowledge of exoplanet statistics. Considering the state of the field $\sim$ 20 years ago, it is reasonable to assume that an observer using this method to interpret observations $\sim 20$ years from now will base their judgment on better prior assumptions than are currently available. Here we speculate on ways in which the prior inputs to our method could be refined in the coming decades.

\subsubsection{What is the largest potentially habitable planet?}
In our work we assume that planets larger than $\sim 1.6 \, R_\oplus$  will form and maintain volatile envelopes over Gyr timescales, thereby resembling ``mini-Neptunes'' more than ``super-Earths''. The most compelling evidence for this comes from density measurements for a limited sample of small planets \citep{Weiss2014,Rogers2015,Chen2017} and the gap in \emph{Kepler} radius occurrence rates near $\sim 1.7\, R_\oplus$ \citep{Owen2013,Fulton2017}, but the exact value of this transition radius and its dependence on parameters such as insolation and spectral type requires further research \citep[e.g.,][]{Fulton2018,MacDonald2019,Martinez2019}. In the near future, TESS \citep{Ricker2014} will detect hundreds of small planets orbiting bright stars \citep{Sullivan2015,Bouma2017,Barclay2018,Ballard2019}. In combination with precise radii measurements by CHEOPS \citep{Broeg2013} and ground-based radial velocity measurements, these discoveries will enhance the sample of $1 - 2\, R_\oplus$ planets with measured densities. Later on, PLATO \citep{PLATO2017} will detect several hundred planets smaller than $2 \, R_\oplus$ on orbits as wide as the Earth's, providing statistics for planet radii over a broader range of insolations than \emph{Kepler}. Models of atmospheric evolution can be used to combine these lines of evidence into a more comprehensive understanding of which planets should have non-habitable volatile envelopes \citep[e.g.,][]{Owen2013,Lopez2014,Gupta2019}.

\subsubsection{What is the period distribution for planets on wide orbits?}
This work relies on extrapolation from \emph{Kepler} occurrence rates for shorter period planets, but ice giants in the habitable zone and beyond could be a significant source of false positives for EEC detection. WFIRST could detect hundreds of wide-orbit planets through its microlensing survey, some with masses lower than the Earth \citep{Barry2011,Penny2019}. PLATO will also detect a number of transiting planets on orbits as wide as the Earth's \citep{PLATO2017}. Even if these data do not fully probe the relevant range of planet radii and periods, they may provide enough points for interpolation to accurately predict the occurrence of terrestrial planets and ice giants within the habitable zone and beyond.

\subsubsection{What do the spectra of false positives look like?}
Our assumed distribution of planet albedos is the least well-vetted prior assumption in this work, but in Cases 6 and 7 we show that prior knowledge of planet albedos could greatly enhance survey efficiency. The most promising avenue for constraining planet albedos in the next decade is through direct imaging of super-Earths and sub-Neptunes in the habitable zone and on wider orbits with ELTs and WFIRST \citep[e.g.,][]{Kasper2010,Traub2016,Savransky2016,Artigau2018,Weinberger2018,Akeson2019}. These observations would be valuable for determining the optimum wavelength ranges for discriminating between ice giants and other types of planets.

Several authors have shown that post-runaway atmospheres like Venus' could develop on extrasolar planets comparable in size to the Earth within the inner edge of the habitable zone, which itself depends on the spectral type of the host star and planetary factors such as mass, rotation rate, and ocean coverage \citep[e.g.,][]{Kopparapu2013,Kopparapu2014,Yang2014,Kodama2018}. Transit spectroscopy with JWST could reveal the atmospheric composition of a small number of terrestrial planets within the runaway greenhouse limit of low-mass stars \citep{Lustig-Yaeger2019}. While not directly measuring the albedo, these observations could reveal whether Venus analogs are common, and modeling efforts could reconstruct their likely appearance in reflected light.

\subsubsection{Testing priors against new data}
Finally, it will be possible for observers to validate their priors during the course of the direct imaging survey. As a simple example, if more faint planets are discovered than predicted under the assumed priors, then it is likely that either the occurrence rates under-predict at small radii or there are more low-albedo planets than are present in the solar system. This information could then be forwarded into a probabilistic analysis of which planets are most likely to be Earth-like before substantial time is committed for follow-up observations.

\section{Conclusions}
We have developed a Bayesian framework with which to infer the properties of a directly imaged planet on the basis of limited photometric data, with the primary goal of identifying exo-Earth candidates for deeper spectroscopic follow-up. This framework is dependent on a multitude of priors drawn from observed exoplanet statistics and a few theoretical models. We use it to characterize the ability of future direct imaging missions to identify potentially habitable planets upon their initial detection using only photometry. We determine the key ambiguities involved in this determination and explore potential solutions, such as constraining the orbit through multiple revisits.

Assuming a uniform prior on a monochromatic albedo, we have found that the detection data alone is not sufficient to determine whether the planet has a potentially habitable size or orbit. In the best cases, a few exo-Earth candidates could be identified with $\sim 50\%$ confidence, but the average EEC would only be identified with $\sim 25\%$ confidence. This translates to a potential false discovery rate of $\sim 75\%$, consistent with previous results \citep{Guimond2018}.

Constraints on the planet's orbit could be achieved through a precursor RV survey or by revisiting the system multiple times. This would allow the observer to constrain the planet to the habitable zone with confidence, but would not eliminate the problem of false positives posed by very small or large planets in the habitable zone. A mass measurement could be somewhat useful for ruling these false positives out, but would by no means be definitive. By revisiting a system multiple times to establish its orbit and measuring the mass astrometrically, an observer could still only distinguish EECs from false positives with a typical confidence of $\sim 65\%$ - but could also identify several individual EECs with high confidence (>90\%).

The use of color information to discriminate between EECs and false positives could dramatically reduce these ambiguities, but requires that prior assumptions be made about the possible appearance of planets (e.g., as a function of their size and insolation). While current data and models do not allow such prior assumptions, in principle this could change over the next decade. In our example, we find that by revisiting a planet to establish its orbit and measuring its brightness in three bands, the strong majority of EECs could be identified with confidence.

Even though we are not always able to confidently identify EECs using our method, we show that a target prioritization strategy which leverages a Bayesian approach will be more efficient than a non-probabilistic approach with the same data. Such an approach could reduce the time required for preliminary spectroscopic follow-up by a factor of two.

Our confidence in these results is dependent on our confidence in the priors we have chosen to use, and some of these - such as the radius and semi-major axis distribution of temperate planets around G dwarfs - are based on extrapolation. Many could be improved in the coming decade - such as the relationship between planet radius, mass, and bulk composition through transit and radial velocity observations and advances in planet formation theory, or the reflected spectra of hot and cold sub-Neptunes with JWST, WFIRST, and ELTs. Near-term efforts to improve our prior knowledge will enable future observers to more efficiently survey nearby systems in search of potentially habitable worlds.

\acknowledgements
The authors are grateful to Christopher Stark for providing a sample of host stars from the LUVOIR yield estimates, and Ilaria Pascucci for offering feedback on the manuscript. A.B. acknowledges support from the NASA Earth and Space Science Fellowship Program under grant No. 80NSS\-C17K0470. The results reported herein benefited from collaborations and/or information exchange within NASA's Nexus for Exoplanet System Science (NExSS) research coordination network sponsored by NASA's Science Mission Directorate. This research has made use of NASA's Astrophysics Data System.

\software{Matplotlib \citep{Hunter07}, NumPy \citep{Oliphant06}, and SciPy \citep{Jones01}.}

\bibliographystyle{aasjournal}
\bibliography{references}

\end{document}